\documentclass[journal]{IEEEtran}

\usepackage{graphicx}
\usepackage{algorithm,algpseudocode}
\usepackage{etex}

\usepackage{amssymb}
\usepackage{mathrsfs}
\usepackage{amsmath}

\usepackage[sans]{dsfont}
\usepackage{stmaryrd}
\usepackage{pifont}
\usepackage{wasysym}

\usepackage{url}
\usepackage[usenames]{color}
\usepackage{caption}
\usepackage{subcaption}
\usepackage{slashbox,pict2e}

\begin{document}

\title{Personalized Prediction of Vehicle Energy Consumption based on Participatory Sensing}

\author{Chien-Ming Tseng and Chi-Kin Chau
\thanks{Chien-Ming Tseng and Chi-Kin Chau are with the Department of EECS, Masdar Institute of Science and Technology, UAE (e-mail: \{ctseng, ckchau\}@masdar.ac.ae).}
\thanks{This paper appears in IEEE Transactions on Intelligent Transportation Systems (DOI:10.1109/TITS.2017.2672880).}
}

\maketitle

\begin{abstract}
The advent of abundant on-board sensors and electronic devices in vehicles populates the paradigm of participatory sensing to harness crowd-sourced data gathering for intelligent transportation applications, such as distance-to-empty prediction and eco-routing. While participatory sensing can provide diverse driving data, there lacks a systematic study of effective utilization of the data for personalized prediction. There are considerable challenges on how to interpolate the missing data from a sparse dataset, which often arises from participatory sensing. This paper presents and compares various approaches for personalized vehicle energy consumption prediction, including a blackbox framework that identifies driver/vehicle/environment-dependent factors and a collaborative filtering approach based on matrix factorization. Furthermore, a case study of distance-to-empty prediction for electric vehicles by participatory sensing data is conducted and evaluated empirically, which shows that our approaches can significantly improve the prediction accuracy.
\end{abstract}


\IEEEpeerreviewmaketitle

\section{Introduction} \label{sec:intro}

Participatory sensing is an emerging paradigm of crowd-sourced data collection and knowledge discovery, which has been applied in diverse applications of pervasive and mobile computing systems \cite{campbell2008parsense}. The basic concept is that a group of users contribute their personal data (possibly, voluntarily) to a third-party data repository, in exchange for the useful knowledge extracted from the collective data, which is then incorporated in personalized applications of individual users.

Vehicles are becoming a vital platform for participatory sensing. First, there are extensive deployments of on-board sensors and in-vehicle information systems, equipped with network connectivity and computing power, acting as effective information collection systems. Second, the wide availability of electronic devices and smartphones carried by passengers can extend the computing and sensing abilities of vehicles. Third, there are abundant off-the-shelf and after-market automotive accessories for gathering driving data and vehicle information. Notably, participatory sensing has been applied in several existing intelligent transportation applications (e.g., traffic status updates in Google Map and Waze).

Furthermore, new intelligent transportation applications can be enhanced by participatory sensing. One of the critical applications is the prediction of {\em distance-to-empty} (DTE) - the distance an electric vehicle (EV) or internal-combustion-engine (ICE) vehicle can reach before its energy/fuel is exhausted. DTE is determined by a variety of factors, such as driving behavior, terrain, types of road, traffic, and vehicle specifications. The conventional approach of DTE prediction employed by car manufacturers is based on the projection of past average vehicle energy efficiency of individual drivers. Such an approach is often perceived to be inaccurate. However, if there is further knowledge about the vehicle, driving behavior and the route to travel, future energy efficiency can be estimated with higher accuracy.

\begin{figure}[!htb]
\includegraphics[width=0.5\textwidth]{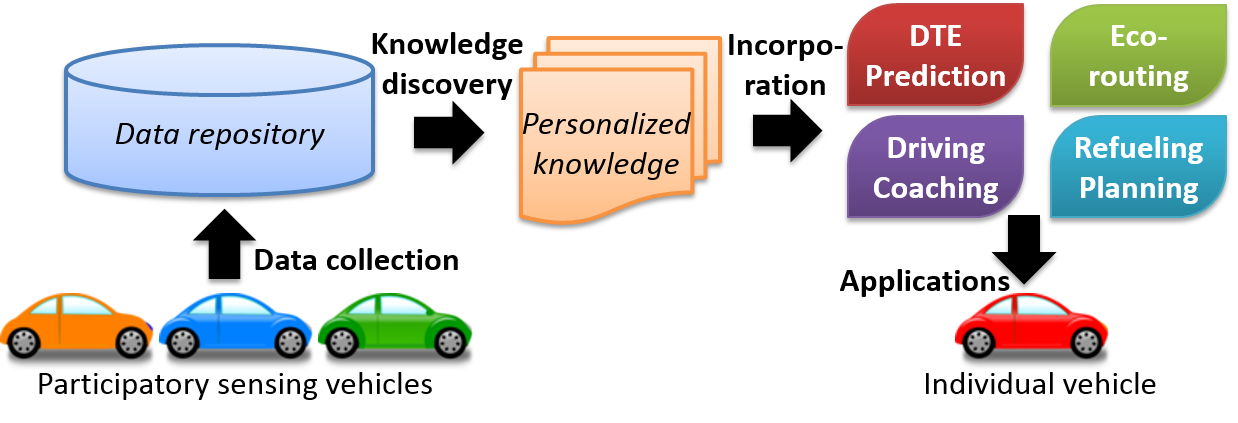} 
\caption{An integrated platform of participatory sensing for personalized applications.}
\label{fig:system}
\end{figure}

The availability of participatory sensing data is able to improve the accuracy of DTE prediction by exploiting the historical data from other drivers. Conceptually, one can identify the characteristics pertaining to specific driver, vehicle or environment. Then, one can harness the measurements from similar drivers, vehicles or environments to assist the prediction. In particular, there are several areas of applications:
\begin{enumerate}

\item {\em Vehicle Centric Applications}: Range anxiety is critical for EVs. Since there are far more ICE vehicles on the road than EVs, one can harness the data collected from ICE vehicles to improve the DTE prediction for EVs.

\item {\em Driver Centric Applications}: With diverse data collected from various drivers, one can compare the driving behavior among drivers. Hence, one can classify driving behavior and provide driving recommendations. 

\item {\em Environment Centric Applications}: Eco-routing or green telematics can be provided by comparing different routes according to energy/fuel consumption prediction. 
    
\end{enumerate}
This paper studies a framework of participatory sensing with an integrated platform of appropriate knowledge discovery and incorporation mechanisms for personalized vehicle/driver/environment centric applications (depicted in Fig.~\ref{fig:system}).

While participatory sensing can provide diverse driving data, there are considerable challenges of harnessing participatory sensing. First, participatory sensing dataset is often sparse and skewed, which does not cover sufficient combinations of vehicle/driver/environment. This calls for an effective approach to interpolate the missing data from a sparse dataset. Second, the dimensionality of dataset may be large due to different combinations of various drivers, vehicles and environments. To enable data analytics, an efficient method is desirable to extract the correlations within data.

This paper explores several approaches of utilizing participatory sensing data for personalized applications:
\begin{enumerate}

\item {\em Comparison with the Average}: One can obtain the average data values (e.g., average speed, stopping duration) from a large dataset for a specific environment. Then the deviation of individual drivers is compensated from the average data values in personalized applications.

\item {\em Collaborative Filtering}: A domain-free data mining technique is to analyze the relationships and interdependencies within a dataset and to identify a smaller set of latent factors that can characterize the observed data. Based on latent factors, one can interpolate the missing data in the dataset. In particular, matrix factorization is a popular solution to realize collaborative filtering.

\item {\em Similarity Matching}: Using a known model of vehicle energy consumption, one can compare the participatory sensing data and find the most similar instances from the available data to estimate the required values.

\end{enumerate}

The notion of average data values has been utilized in previous papers \cite{ganti2010greengps,stefan2013modularecev}, which has a disadvantage of requiring a large dataset. Collaborative filtering \cite{gemulla2011lgmfgd,yehuda2009mftrc} is a general technique in data mining without leveraging the detailed knowledge of underlying model. However, the presence of specific knowledge in vehicle energy consumption can potentially improve the accuracy and effectiveness.

This paper presents several viable approaches of utilizing participatory sensing data for personalized prediction of vehicle energy consumption. In particular, a blackbox framework is presented to effectively identify driver/vehicle/environment dependent factors from participatory sensing data for personalized prediction. To demonstrate the effectiveness of our approaches, a case study of distance-to-empty prediction for EVs based on the participatory sensing data is conducted and validated empirically, which shows that our approaches can significantly improve the prediction accuracy.

{\bf Outline of Paper:} The related work is first presented in Sec.~\ref{sec:related}. The methodologies of personalized prediction approaches are presented in Secs.~\ref{sec:method}-\ref{sec:drvselect}. The empirical evaluations of our approaches are given in Sec.~\ref{sec:expval}. A case study that utilizes our results is discussed in Sec.~\ref{sec:case}.

\section{Related Work} \label{sec:related}

\subsection{Vehicle Energy Consumption Models}

Modeling vehicle energy consumption has been the subject of a number of research papers. One popular method is the model-based approach, which is based on vehicle dynamics to model the consumption behavior of ICE vehicles \cite{karin2012eeroutealg} and EVs \cite{eugene2013rtbattery}. Energy consumption estimation can use a blackbox approach. For example, a statistical approach using regression model to estimate the energy consumption of ICE vehicles is presented in \cite{alesaandn2002statmdlfc}. The energy consumption rate of different vehicles and roads can also be clustered to characterize the energy consumption of general vehicle and road types \cite{stefan2013modularecev}.

\subsection{Data Collection of Vehicle Energy Consumption}
The accuracy of energy consumption prediction can be enhanced by collecting more information. Two crucial factors of energy consumption prediction are the future speed profiles and future environmental factors (e.g., temperature, wind speed or route grade), which may be highly dynamic and difficult to predict.
One method to estimate the future speed profiles is to utilize Markov chain \cite{javier2013mbasedrdr}.
Also, one can deploy sensor networks, by which stationary measurements at specific locations, such as traffic, average speed, speed limit and route grade can be measured. There are a number of papers focusing on utilizing such information \cite{stefan2013modularecev}. However, the traffic data in these papers is usually static, which may have a large deviation in dynamic traffic. A study that integrates the real-time traffic sensor data to predict the energy consumption and emission of ICE vehicles is presented in \cite{qichiyang2011arterialmdltraj}.
One can also obtain the estimated information from social networks and participatory sensing. Participatory sensing can provide mobile measurements and good geographic penetrations \cite{cloudthink15}. For example, \cite{dornbush2007streetsmart} shows that the estimation of stochastic effects which impact the travel velocity and acceleration profiles can be crowd-sourced to identify traffic congestion. 
Our previous work employs participatory sensing for DTE prediction \cite{cmtseng2015pardte,cmtseng2014social}, which is extended in this paper.

\subsection{Applications of Vehicle Energy Consumption}
The integration of energy consumption prediction and data collection enables many applications. One application is the estimation of DTE, based on the prediction of vehicle energy efficiency (i.e., energy intensity), which is employed in production vehicles \cite{burke1994electronic}. DTE can be estimated by measuring the mean energy consumption over short and long distances \cite{rodgers2013conventional}. To account for the deviation between the historical and future energy intensity, a regression model can be used to predict the future energy intensity given future route information \cite{anastasia2014onpevreg}. Route features from sensor data can be clustered to identify the driving pattern for EV range estimation \cite{haiyu2012drvpatternev}.
Another application is eco-routing. A system based on road characteristics and current prevailing traffic conditions is presented in \cite{kanok2012ecorttraffic}. It can provide users a more economic and safer route with reasonable speed instead of only driving at a lower speed. Another study utilizing average participatory sensing data and static traffic information for eco-routing system of ICE vehicles is presented in \cite{ganti2010greengps}.
Route-type based energy consumption prediction can be implemented using OpenStreetMap (OSM) data. There are some studies using the OSM data to predict the EV range \cite{martin2011eoptev}. A cloud based prediction system considers the deviation between the mean energy consumption and that of different condition (e.g., traffic congestion or driving behavior) in \cite{stefan2013syscloud}. The route-type based energy consumption model requires a complete map database including speed limit, route type and traffic information, which may not be available everywhere.

\medskip

This work differentiates from the previous work in several aspects: (1) We compare various personalized prediction approaches of vehicle energy consumption. (2) We present a novel blackbox framework to extract driver/vehicle/ environment-dependent factors. (3) We conduct a case study of DTE prediction for EVs using different approaches. 

\section{Methodology and Background} \label{sec:method}

This section presents the relevant methodology and background related to vehicle energy consumption models.

\subsection{Areas of Factors}

While there are many factors to determine vehicle energy consumption, they can be classified by three broad areas:
\begin{itemize}

\item {\bf Driver}:
The driver who controls the vehicle has a direct impact on the vehicle movement. Different drivers exhibit different preferences for stop/start and acceleration, aggression in various scenarios, propensity for hypermiling, etc. Psychological and behavioral traits of drivers also affect vehicle energy efficiency.

\item {\bf Vehicle}:
Different types of vehicles consume energy differently. ICE vehicles are characterized by the engine types and gear shifts, whereas hybrid and EVs are affected by battery performance and regenerative braking. The sizes and weights of vehicles often determine the efficiency of kinetic energy conversion, so SUVs and trucks are usually less energy-efficient than sedan and compact vehicles.

\item {\bf Environment}:
{\color{black}
The environmental factors include traffic and roads. Traffic for a road segment depends on a plethora of factors, including time-of-day, day-of-year, special events, which may follow a certain pattern. The types of roads also affect drivers' behavior differently, which can be divided into three main categories: small public or private roads with urban traffic, lower capacity ``urban'' highways, and higher capacity freeways. Other environmental factors, such as road grades and weather types, can also be considered.
}
\end{itemize}

The historical data of vehicle speed profiles can be identified by a combination of (driver, vehicle, environment), referred as a {\em data point}. This paper aims to predict the energy consumption for each data point. Through participatory sensing, a dataset of measured energy consumption for a relatively small number of data points are collected. This paper addresses the challenge of data interpolation with good accuracy.

\subsection{Types of Models}

There are two main types of energy consumption models:
\begin{itemize}

\item {\bf White-box Model}: A straightforward approach is to employ a white-box microscopic behavior model of each vehicle that comprehensively characterizes the engine performance, vehicle mechanics, battery systems, etc. To incorporate traffic information, one can rely on a macroscopic traffic database collected from a network of loop sensors along specific road segments. However, such a white-box vehicle model requires a large amount of data for calibration and detailed knowledge specific to a particular vehicle. Also, the availability and access of accurate traffic information is often limited to certain authorized parties only.

\item {\bf Blackbox Model}: A blackbox approach is more desirable that requires minimal knowledge of vehicle model with only a small set of measurable variables and parameters. The variables and parameters depend on the combinations of (driver, vehicle, environment). In the subsequent sections, the variables and parameters obtained in the blackbox model will be utilized for collaborative filtering and similarity matching.

\end{itemize}

\subsection{Energy Consumption Model}

This section describes a linear blackbox model of vehicle energy consumption that has been used extensively in the literature \cite{ganti2010greengps,stefan2013modularecev,alesaandn2002statmdlfc,cmtseng2015pardte,cmtseng2016privacy}.
Denote a driver by ${\sf D}$, a vehicle model by ${\sf V}$, and a particular environment (e.g., a segment of route and time-of-day) by ${\sf R}$. Each energy consumption is represented by a numerical value ${E}_{{\sf D},{\sf V},{\sf R}}$, indexed by the tuple $({\sf D},{\sf V},{\sf R})$. All the entries of energy consumption values form a $3$-dimensional tensor, denoted by $[{E}_{{\sf D},{\sf V},{\sf R}}]$.

While there are sophisticated approaches of estimating the moving vehicle energy consumption by white-box microscopic behavior models \cite{eugene2013rtbattery,karin2012eeroutealg,javier2013mbasedrdr,haiyu2012drvpatternev,ckchau2016phevopt,ckt2016dm}, these models are rather difficult to implement. Many parameters are required, for example, engine efficiency, transmission efficiency, regenerative braking efficiency, etc. However, in practice, these parameters are hard to obtain. Therefore, this paper utilizes a blackbox approach without the detailed knowledge of vehicle mechanics. This approach maximizes the applicability for a wide range of scenarios arising from participatory sensing.

The total energy consumption ${E}$ of driver {\sf D} with vehicle model {\sf V} in a particular environment {\sf R} is given by:
\begin{equation}
{E}_{{\sf D},{\sf V},{\sf R}} = {E}^{\rm mv}_{{\sf D},{\sf V},{\sf R}} + {E}^{\rm id}_{{\sf D},{\sf V},{\sf R}}  \label{eq:totalmodel}
\end{equation}
where ${E}^{\rm mv}_{{\sf D},{\sf V},{\sf R}}$ is the moving vehicle energy consumption and ${E}^{\rm id}_{{\sf D},{\sf V},{\sf R}}$ is the idle vehicle energy consumption.

\medskip

\begin{figure*}[!htb]
    \begin{subfigure}[!htb]{0.33\textwidth}
        \centering
        \includegraphics[width=\textwidth]{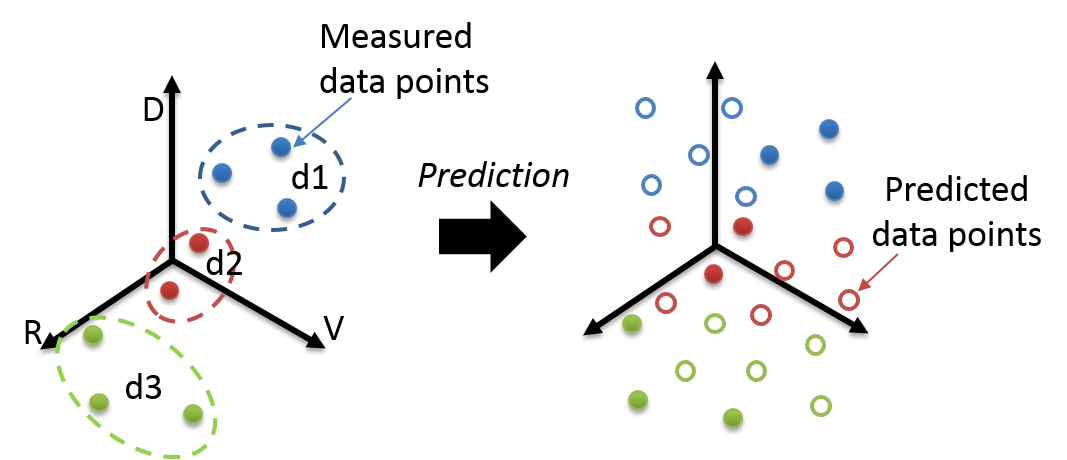}
        \caption{Measured and interpolated data points in a sparse dataset.}
        \label{fig:dataspace}
    \end{subfigure}%
    ~
    \begin{subfigure}[!htb]{0.28\textwidth}
        \centering
        \includegraphics[width=\textwidth]{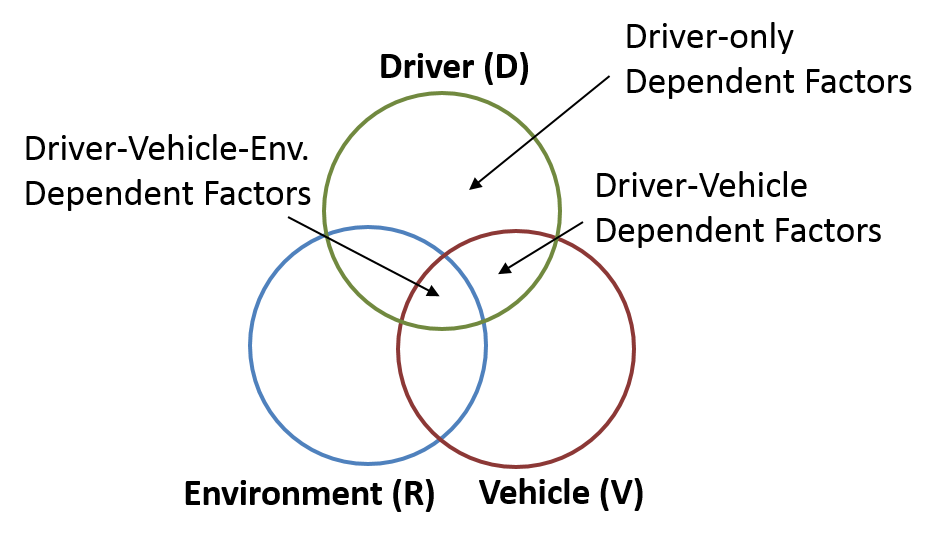}
        \caption{Factors and dependence.}
        \label{fig:featurespace}
    \end{subfigure}
    ~
    \begin{subfigure}[!htb]{0.36\textwidth}
        \centering
        \includegraphics[width=\textwidth]{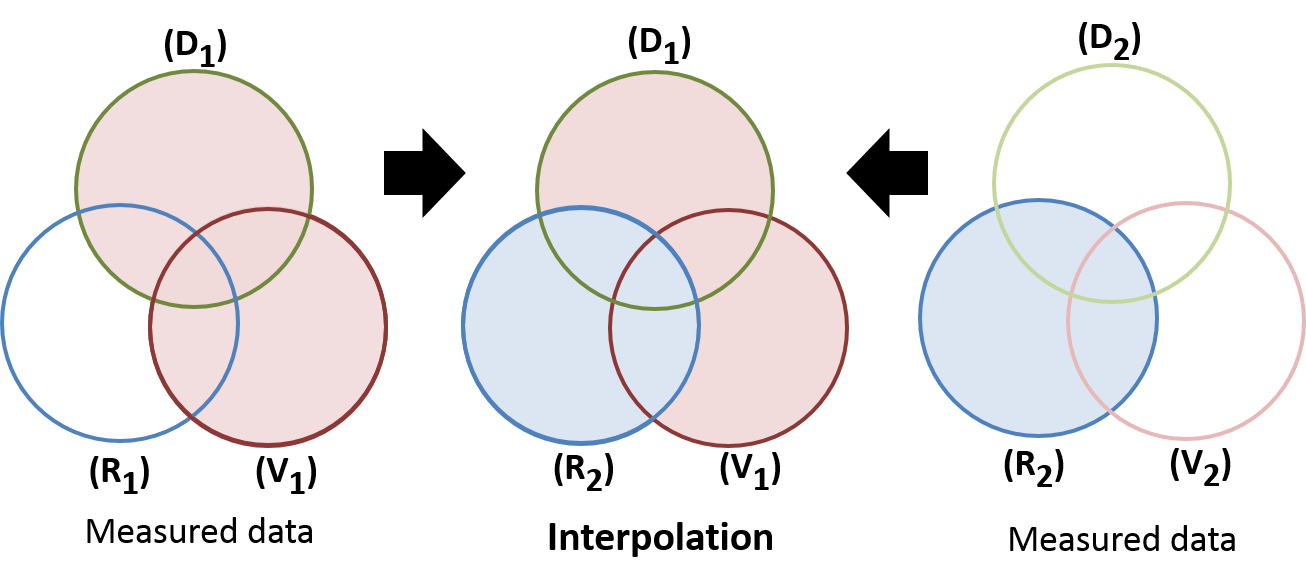}
        \caption{Interpolation of data points by substitution of factors from the most similar measured data points.}
        \label{fig:prediction}
    \end{subfigure}

	\caption{Illustrations for interpolation of missing data points.}
\end{figure*}

\subsubsection{Moving Vehicle Energy Consumption}

With respect to a particular combination of $({\sf D},{\sf V},{\sf R})$, the moving vehicle energy consumption ${E}^{\rm mv}$ has unit in liter or kWh. Next, the subscript $_{{\sf D},{\sf V},{\sf R}}$ is dropped for brevity.

In this paper, ${E}^{\rm mv}$ (denoted by $\hat{E}^{\rm mv}$) is estimated by a linear equation of several measurable variables from vehicles\footnote{Some of the variables are selected based on \cite{eva2001idrvpattern}, which analyzed more than 20 thousand data points from 45 drivers to identify the most significant factors of fuel consumption and emission.}:
\begin{align}
& \hat{E}^{\rm mv} = \notag \\
& \hspace{-10pt} \begin{bmatrix}
\alpha_{v,1}\\
\alpha_{v,2}\\
\vdots \\
\alpha_{v,r}
\end{bmatrix}^{T}
\begin{bmatrix}
{v}\\
{v}^2\\
\vdots \\
{v}^r
\end{bmatrix}+\hspace{-4pt}
\begin{bmatrix}
\vec{\alpha}_{d,1}\\
\vec{\alpha}_{d,2}\\
\vdots \\
\vec{\alpha}_{d,k}
\end{bmatrix}^{T}
\begin{bmatrix}
\vec{{d}}\\
{\vec{d}}^2\\
\vdots \\
{\vec{d}}^k
\end{bmatrix}+\hspace{-4pt}
\begin{bmatrix}
\vec{\alpha}_{a,1}\\
\vec{\alpha}_{a,2}\\
\vdots \\
\vec{\alpha}_{a,m}
\end{bmatrix}^{T}
\begin{bmatrix}
\vec{{a}} \\
{\vec{a}}^2 \\
\vdots \\
{\vec{a}}^m
\end{bmatrix}+\hspace{-4pt}
\begin{bmatrix}
\alpha_g\\
\alpha_{\ell}\\
c
\end{bmatrix}^{T}
\begin{bmatrix}
{g} \\
{\ell} \\
1
\end{bmatrix} \label{eqn:movmodel}
\end{align}
where
\begin{itemize}

\item ${v}$ is the continuous average speed (i.e., the average speed without idling). The higher powers\footnote{Sec.~\ref{sec:expval} will empirically determine the proper powers of parameters.} of ${v}$ like ${v}^2,...,{v}^r$ are also considered.

\item $\vec{{d}} = (\tau_{d}, \mu_{d}, \sigma_{d})$ is the deceleration tuple:

\begin{itemize}

\item $\tau_{d}$ is the total duration of deceleration.

\item $\mu_{d}$ is the mean deceleration (i.e., the sum of deceleration values divided by the deceleration duration).

\item $\sigma_{d}$ is the standard deviation of deceleration.

\end{itemize}
Denote the higher powers of components in the deceleration tuple by $\vec{{d}}^k = (\tau_{d}^k, \mu_{d}^k, \sigma_{d}^k)$.

\item $\vec{{a}}$ is the acceleration tuple (similar to $\vec{{d}}$).

\item ${g}$ is the mean absolute value of gyroscope along the moving direction.

\item ${\ell}$ is the auxiliary load of idling, which is the baseline measurement when the vehicle is not moving.

\item ${c}$ is a normalization constant.

\item ${\alpha}_{v} \triangleq ({\alpha}_{v,1},..., {\alpha}_{v,r})$, ${\alpha}_{d} \triangleq ({\alpha}_{d,1},..., {\alpha}_{d,k})$, ${\alpha}_{a} \triangleq ({\alpha}_{a,1},$ $..., {\alpha}_{a,k}), \alpha_g, \alpha_\ell$ are the corresponding coefficients.

\end{itemize}
Note that the coefficients ${\alpha}_{d} \triangleq ({\alpha}_{d,1},..., {\alpha}_{d,k})$ can effectively capture the regenerated energy of EVs.

\medskip

\subsubsection{Idle Vehicle Energy Consumption}
\

Similarly, a blackbox approach is used to estimate the idle vehicle energy consumption. The subscript $_{{\sf D},{\sf V},{\sf R}}$ is dropped for brevity.
With respect to a particular combination of $({\sf D},{\sf V},{\sf R})$, the idle vehicle energy consumption ${E}^{\rm id}$ (denoted by $\hat{E}^{\rm id}$) is estimated by a linear equation:

\begin{equation}
\hat{E}^{\rm id} = \beta_1 \mu {\ell} + \beta_2 \omega \label{eqn:idlmodel} 
\end{equation}

where
\begin{itemize}

\item $\mu$ is the total idle duration.

\item ${\ell}$ is the auxiliary load of idling.

\item $\omega$ is the outdoor temperature.

\item $\beta_1, \beta_2$ are the coefficients.

\end{itemize}

The parameters $v, \ell$ can be obtained from standard OBD data inquiry from vehicles, whereas $\vec{{d}}, \vec{{a}}, \mu$ can be computed from speed profiles, $g$ can be obtained from smartphones, and $\omega$ can be obtained from online weather data.

\subsection{Estimation of Coefficients}

The coefficients (${\alpha}_{v}, \vec{\alpha}_{d}, \vec{\alpha}_{a}, \alpha_g, \alpha_\ell, c, \beta_1, \beta_2$) in Eqns. (\ref{eqn:movmodel})-(\ref{eqn:idlmodel}) can be estimated by the standard regression method, if sufficient measured data (${v}, \vec{{d}}, \vec{{a}}, g, \ell, \mu, \omega$) and the respective energy consumption data ($\hat{E}^{\rm mv}, \hat{E}^{\rm id}$) are provided.
Assume that each driver-vehicle pair $({\sf D},{\sf V})$ has collected sufficient historical personal driving data, and hence, the coefficients can be estimated for the respective environment ${\sf R}$.
One notable advantage of regression method is that it is less susceptible to random noise, which can arise from various sources (e.g., due to time synchronization in data sampling, mechanic damping, inaccurate measurements).

\section{Interpolating Participatory Sensing Data} \label{sec:drvselect}

Given a dataset of driving data collected by participatory sensing, a data point can be visualized as a point in a $3$-dimensional Euclidean space, indexed by $({\sf D},{\sf V},{\sf R})$. The participatory sensing dataset is usually sparse, consisting of a skewed and clustered distribution of data points. In order to predict the vehicle energy consumption for the data points that are not collected from participatory sensing, we seek to interpolate the missing data points to cover the space of dataset. An illustration is depicted in Fig.~\ref{fig:dataspace}.

Next, three major data interpolation approaches by similarity matching, matrix factorization, and comparison with the average are presented.

\subsection{Similarity Matching}

Similarity matching is related to neighborhood-based collaborative filtering.
The three areas of factors (i.e., driver, vehicle, and environment) that determine the vehicle energy consumption are not necessarily exclusive. There are factors that can belong to multiple aspects. For example, the speed of a vehicle depends on both driver and environment. Abstractly, the factors can be visualized by a Venn diagram (see Fig.~\ref{fig:featurespace}).

After characterizing the factors and their dependence, the interpolation of missing data points can be attained by suitable substitution of factors from the most similar measured data points. For example, see Fig.~\ref{fig:prediction} for an illustration. After obtaining the measured data for $({\sf D}_1,{\sf V}_1,{\sf R}_1)$ and $({\sf D}_2,{\sf V}_2,{\sf R}_2)$, we aim to estimate the energy consumption for $({\sf D}_1,{\sf V}_1,{\sf R}_2)$. If ${\sf D}_1$ is similar to ${\sf D}_2$ and ${\sf V}_1$ is similar to ${\sf V}_2$, then one can replace the factors that depend on ${\sf R}_1$ in $({\sf D}_1,{\sf V}_1,{\sf R}_1)$ by those depend on ${\sf R}_2$ in $({\sf D}_2,{\sf V}_2,{\sf R}_2)$.

The energy consumption model in Eqns.~(\ref{eqn:movmodel})-(\ref{eqn:idlmodel}) provides a convenient way to extract the factors of driver, vehicle, and environment dependence. In Table~\ref{tab:depend}, the dependence of each parameter is heuristically assigned based on the major observable impacts from the driver, vehicle or environment.

\begin{table}[htb!] \centering
{
\begin{tabular}{@{}c@{\ }|@{\ }c@{\ }|@{\ }c@{\ }|@{\ }c@{}}
  \hline  \hline
    \ & Driver- \ & Vehicle- \ & Environment- \\
    \ & dependent \ & dependent \ & dependent \\
  \hline
   ${v}, \vec{{d}}, \vec{{a}}, g$, $\ell$& $\checkmark$ &  & $\checkmark$ \\
   $\mu, \omega$ & & & $\checkmark$ \\
  \hline
   $\alpha_v, \vec{\alpha}_d, \vec{\alpha}_a, \alpha_g$ & & $\checkmark$ &  \\
   $\alpha_{\ell}$, $c$ & & $\checkmark$ & \\
   $\beta_1,\beta_2$ & $\checkmark$ & $\checkmark$ & \\
  \hline  \hline
\end{tabular} }
\caption{Dependence of parameters and coefficients.} \label{tab:depend}
\end{table}

For the coefficients, it is assumed that their dependence is complementary to that of the respective parameters. For example, the average speed ${v}$ is more likely affected by the driver and environment, while to a less extent by the type of vehicle. Hence, coefficient $\alpha_v$ is considered to be vehicle-dependent, such that the product $\alpha_v {v}$ will be specific to a particular tuple $({\sf D},{\sf V},{\sf R})$. The dependence of coefficients will be empirically validated in Sec.~\ref{sec:expval}.

The interpolation of missing data points can be attained by the substitution of parameters and coefficients in the vehicle energy consumption model. Consider an example in Fig.~\ref{fig:prediction}. Let the parameters and coefficients for $({\sf D1},{\sf V1},{\sf R1})$ be (${\alpha}_{v}, \vec{\alpha}_{d}, \vec{\alpha}_{a}, \alpha_g, \alpha_\ell, c, \beta_1,\beta_2$) and (${v}, \vec{{d}}, \vec{{a}}, g, \ell, \mu, \omega$), and those for $({\sf D2},{\sf V2},{\sf R2})$ be (${\alpha}'_{v}, \vec{\alpha}'_{d}, \vec{\alpha}'_{a}, \alpha'_g, \alpha'_\ell, c', \beta_1', \beta_2'$) and (${v}', \vec{{d}}', \vec{{a}}', g', \ell', \mu', \omega'$). To estimate the energy consumption of $({\sf D1},{\sf V1},{\sf R2})$, (${\alpha}_{v}, \vec{\alpha}_{d}, \vec{\alpha}_{a}, \alpha_g, \alpha_\ell, c, \beta_1, \beta_2$) and (${v}', \vec{{d}}', \vec{{a}}', g', \ell, \mu', \omega $) can be used in the vehicle energy consumption model.

To determine the similarity among drivers and vehicles, two approaches of similarity matching by speed profile matching and driving habit matching are presented next.

\medskip

\subsubsection{Speed Profile Matching}
\

One can characterize the similarity between a pair $({\sf D},{\sf V})$ and $({\sf D}',{\sf V}')$ under the same environment ${\sf R}$ by comparing the respective speed profiles (i.e., the plots of speed against traveled distance). Since speed profiles are time series, {\em dynamic time warping} (DTW)  \cite{Stan2007DTW} can be used as a metric for determining the similarity, and identifying the corresponding similar regions between two time series, which has been applied in many applications (e.g., speech recognition).

The basic idea of DTW is to determine an optimal alignment between two time series. Consider two time series $X = (x[t])_{t=1}^{n_X}$ and $Y = (y[t])_{t=1}^{n_Y}$ of lengths $n_X$ and $n_Y$ respectively.
A {\em warp path} is defined as $W = (w[k])_{k=1}^{n_W}$, where the $k$-th element is $w_k=(i,j)$, such that $i$ is an index from time series $x[i]$ and $j$ is an index from time series $y[j]$. $n_W$ is the length of the warp path $W$, such that $\max(n_X,n_Y)\leq n_W <n_X+n_Y$.
The warp path $W$ is subject to the following constraints:
\begin{enumerate}

\item $w[1] = (1, 1)$ and $w[n_W]= (n_X, n_Y)$;

\item if $w[k] = (i, j)$ and $w[k+1] = ({i'}, {j'})$, then $i \leq {i'} \leq i+1$ and $j \leq {j'} \leq j+1$.

\end{enumerate}
The warp path of minimum distance ${\sf dist}(W^\ast)$ is defined by:
\begin{equation}
{\sf dist}(W^\ast)=\min_{w} \sum_{k=1}^{n_W} d(w[k]) \label{eqn:eqroumod1}
\end{equation}
where each $d(w[k]) = |x[i]-y[j]|$ is the distance of the coordinates $(i,j)$ of the $k$-th element in $W$.
A simple approach to determine an optimal warp path between two time series is using dynamic programming. But there are other more efficient algorithms with linear running time \cite{Stan2007DTW}.

Suppose each trip is divided into a sequence of segments $({\sf R}^i)$.
Let $v_{{\sf D},{\sf V},{\sf R}^i}[t]$ be the time series of speed profile for tuple $({\sf D},{\sf V},{\sf R}^i)$. For each pair of $({\sf D},{\sf V},{\sf R}^i)$ and $({\sf D'}, {\sf V'}, {\sf R}^i)$, define
\begin{equation}
\chi_{({\sf D},{\sf V}), ({\sf D'}, {\sf V'})}^{{\sf R}^i} \triangleq {\sf dist}(W^\ast)
\end{equation}
where $W^\ast$ is the minimum-distance warp path between the time series $v_{{\sf D},{\sf V},{\sf R}^i}[t]$ and $v_{{\sf D'}, {\sf V'}, {\sf R}^i}[t]$.

Let ${\mathscr R}({\sf D},{\sf V})$ be a set of segments that have speed profiles measured with $({\sf D},{\sf V})$. Namely, if ${\sf R}^i \in {\mathscr R}({\sf D},{\sf V})$, then the speed profile $v_{{\sf D},{\sf V},{\sf R}^i}[t]$ exists in the dataset.
Define a similarity metric $\bar{\chi}_{({\sf D},{\sf V}),({\sf D'},{\sf V'})}$ between each pair of $({\sf D},{\sf V})$ and $({\sf D'},{\sf V'})$ by the average minimum warp path distance over all segments:
\begin{equation}
\bar{\chi}_{({\sf D},{\sf V}),({\sf D'},{\sf V'})} \triangleq
\frac{\displaystyle\sum_{ {\sf R}^i \in {\mathscr R}({\sf D},{\sf V}) \cap {\mathscr R}({\sf D'}, {\sf V'})}
\chi_{({\sf D},{\sf V}),({\sf D'},{\sf V'})}^{{\sf R}^i}
}{|{\mathscr R}({\sf D},{\sf V}) \cap {\mathscr R}({\sf D'}, {\sf V'})|}
    \label{eqn:cormetric}
\end{equation}
Note that $\bar{\chi}_{({\sf D},{\sf V}),({\sf D'},{\sf V'})} = \infty$, if ${\mathscr R}({\sf D},{\sf V}) \cap {\mathscr R}({\sf D'},{\sf V'}) = \varnothing$.

\begin{figure}[!htb]
\center
 \includegraphics[width=0.52\textwidth]{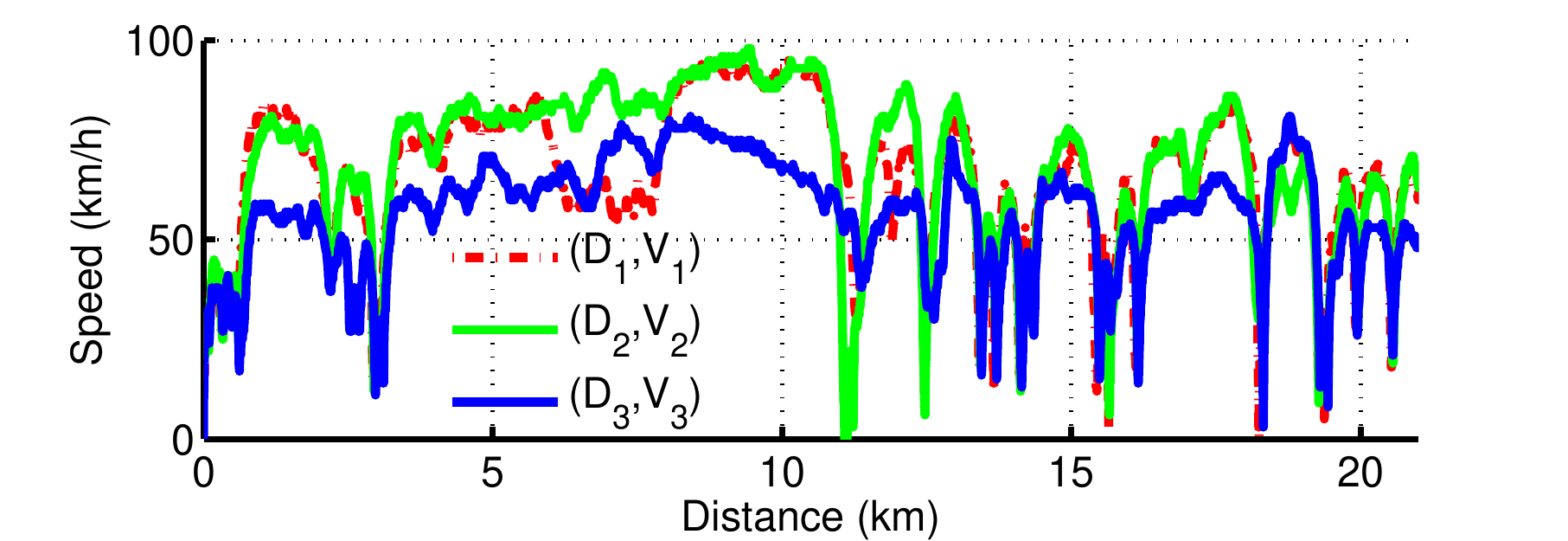}
\caption{{Speed profiles of three drivers on the same trip. $\chi_{({\sf D}_1, {\sf V}_1), ({\sf D}_2, {\sf V}_2)}^{{\sf R}^1} = 1.1385$ and $\chi_{({\sf D}_1, {\sf V}_1), ({\sf D}_3, {\sf V}_3)}^{{\sf R}^1} = 1.3883$. }}
\label{fig:DTWdist}
\end{figure}

For example, the speed profiles of three driver-vehicle pairs $({\sf D}_1,{\sf V}_1), ({\sf D}_2,{\sf V}_2),({\sf D}_3,{\sf V}_3)$ for the same trip of a certain road ${\sf R}^1$ are plotted in Fig.~\ref{fig:DTWdist}. Smaller minimum warp path distance is observed to have closer similarity in the speed profile; namely, $({\sf D}_1,{\sf V}_1)$ is more similar to $({\sf D}_2,{\sf V}_2)$ than $({\sf D}_3,{\sf V}_3)$.

This paper uses $\bar{\chi}_{({\sf D},{\sf V}),({\sf D'},{\sf V'})}$ to characterize the similarity between each pair of $({\sf D},{\sf V})$ and $({\sf D'},{\sf V'})$. The tuple $({\sf D'},{\sf V'})$ with the smallest value $\bar{\chi}_{({\sf D},{\sf V}),({\sf D'},{\sf V'})}$ is identified for estimating energy consumption of $({\sf D},{\sf V})$. For finding multiple similar data points, $k$-nearest neighbors ($k$-NN) clustering is employed to find the $k$ most similar speed profiles with $({\sf D},{\sf V})$.

\medskip

\subsubsection{Driving Habit Matching}
\

Speed profiles are not always available for the same environment. An alternative is to rely on the available data collected from other environments. An important factor for vehicle energy consumption is the acceleration/deceleration \cite{eva2001idrvpattern}. The accelerating behavior of the drivers is related to vehicle energy efficiency. On the other hand, aggressive decelerations, usually inducing rear-end collisions, is related to driving awareness.

\begin{figure}[!htb]
    \centering
        \includegraphics[width=.48\textwidth]{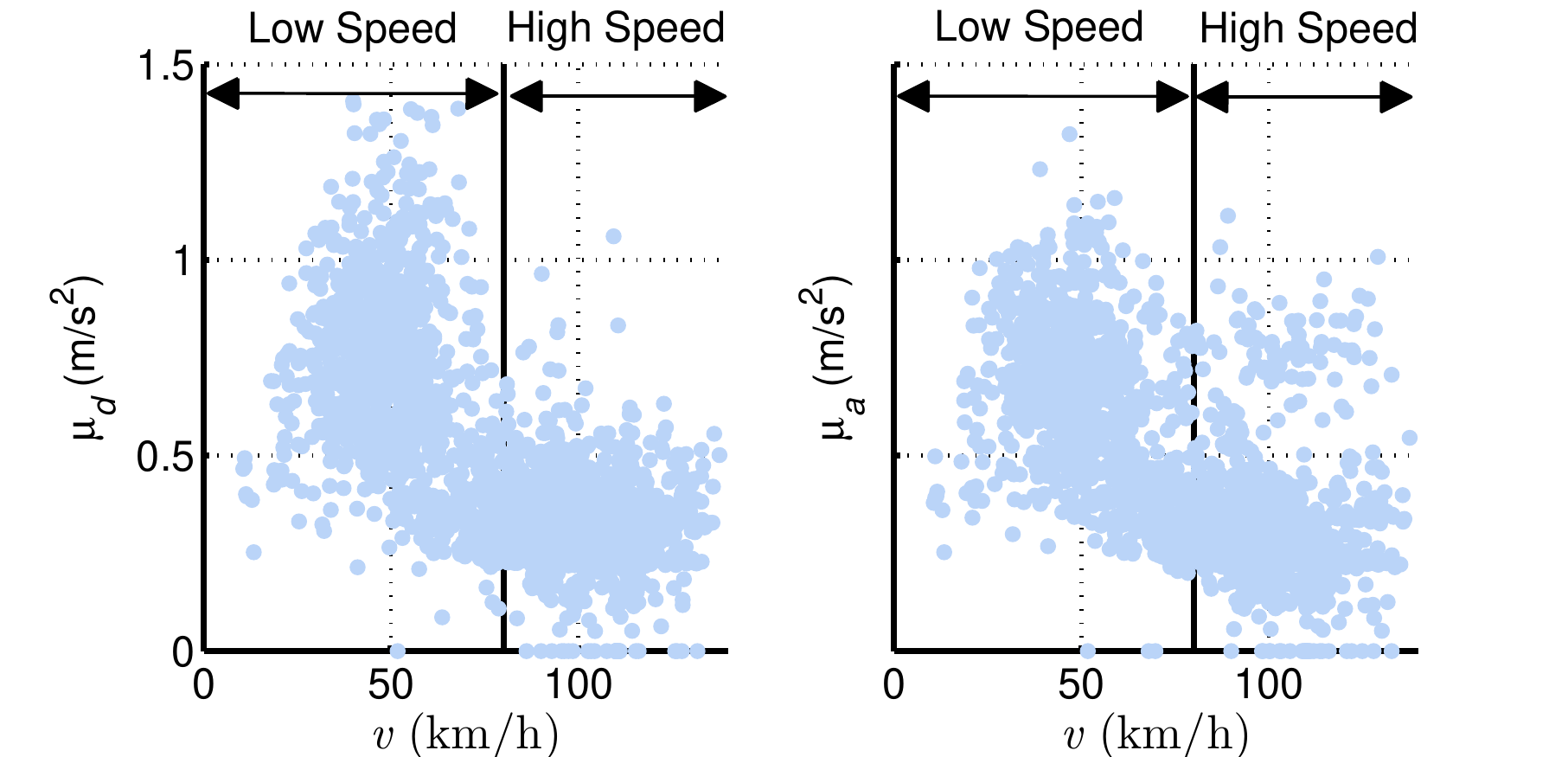}
        \caption{Mean acceleration and deceleration amplitudes vs. continuous average speed of each segment of a driver.}
        \label{fig:accall}
\end{figure}

For example, the mean acceleration and deceleration ($\mu_{d}, \mu_{a}$) against the continuous average speed of each segment from the data of a driver are plotted in Fig.~\ref{fig:accall}. It is observed that the decelerations/accelerations tend to be higher at a low speed (possibly, due to stop-and-go behavior), whereas lower decelerations/accelerations can be found at a high speed. Low accelerations are usually due to cruise control or mindful drivers, while high accelerations are due to aggressive driving.

Therefore, we are motivated to use the average acceleration and deceleration as a metric to characterize driving habits. However, we may not have collected sufficient measurements for every vehicle speed. Hence, we normalize the distribution of data to obtain a better estimation of the average. First, divide the range of vehicle speed into a sequence of intervals with width $\Delta v$ (i.e., $[v, v+\Delta v]$). This paper considers $\Delta v=10$km/h.	
For each interval $[v, v+\Delta v]$, let $\gamma_{a}^{v}({\sf D},{\sf V})$ be the mean value of the acceleration measurements within $[v, v+\Delta v]$.
Define the {\em estimated average acceleration} by the average of the mean values in all intervals by $\bar{\gamma}_{a}({\sf D},{\sf V})$. To avoid bias, we ignore the intervals in which the number of the data points is less than 10.

\begin{figure}[!htb]\hspace{-5pt}
        \includegraphics[width=.5\textwidth]{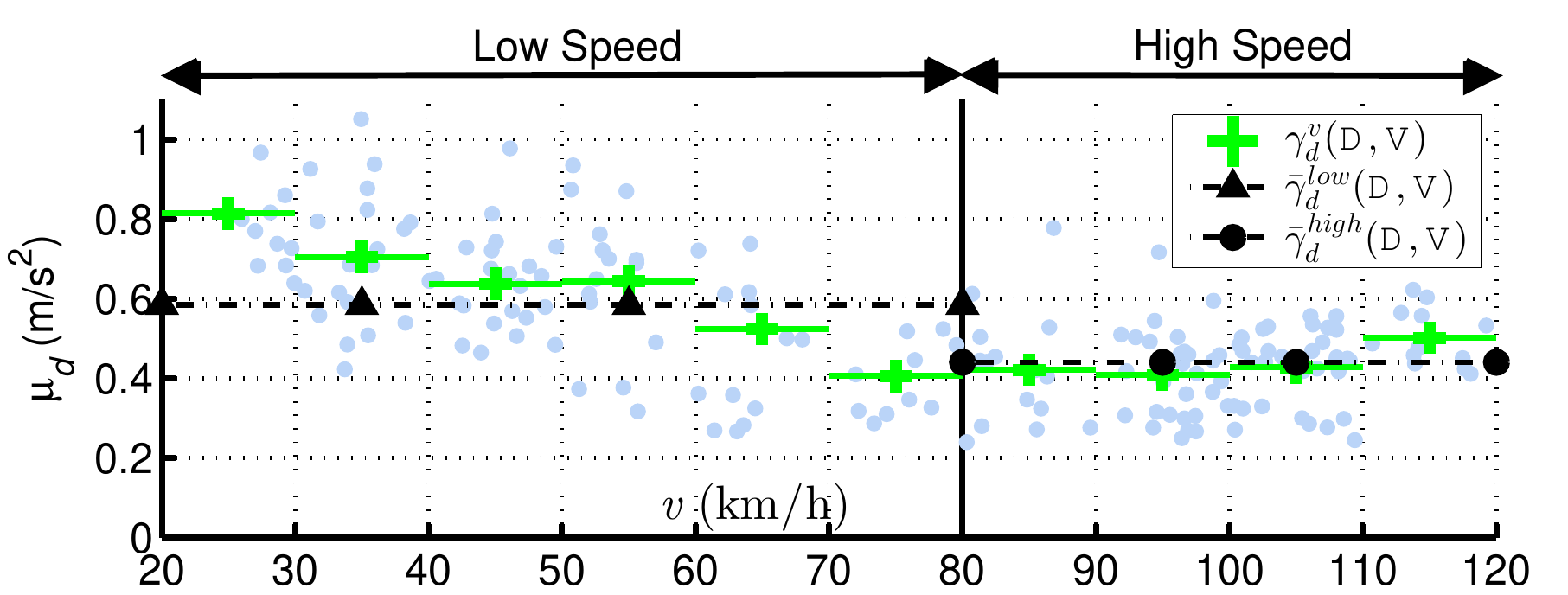}
        \caption{An illustration of $\bar{\gamma}_{d}({\sf D},{\sf V})$ and ${\gamma_{d}^{v}({\sf D},{\sf V})}$.}
        \label{fig:drvbin}
\end{figure}

Because a difference is observed in high-speed and low-speed driving habits, we define different estimated average accelerations for the intervals above or below a threshold $v_{\rm th}$:
\begin{enumerate}
  \item Low-speed estimated average acceleration $\bar{\gamma}_{a}^{\rm low}({\sf D},{\sf V})$.

  \item High-speed estimated average acceleration $\bar{\gamma}_{a}^{\rm high}({\sf D},{\sf V})$.

\end{enumerate}
Similarly, define $\bar{\gamma}_{d}^{\rm low}({\sf D},{\sf V})$ and $\bar{\gamma}_{d}^{\rm high}({\sf D},{\sf V})$ for deceleration.
Fig.~\ref{fig:drvbin} depicts an illustration of ${\gamma}_d^{v}({\sf D},{\sf V})$, $\bar{\gamma}_{d}^{\rm low}({\sf D},{\sf V})$ and $\bar{\gamma}_{d}^{\rm high}({\sf D},{\sf V})$ from a dataset of driving data.

Heuristically, this paper sets $v_{\rm th}=80$km/h, because this speed limit usually sets the difference between highways and suburban roads.
The average deceleration/acceleration tuple $\big(\bar{\gamma}_{a}^{\rm low}({\sf D},{\sf V}), \bar{\gamma}_{a}^{\rm high}({\sf D},{\sf V}), \bar{\gamma}_{d}^{\rm low}({\sf D},{\sf V}), \bar{\gamma}_{d}^{\rm high}({\sf D},{\sf V})\big)$ can capture the driving habit of each $({\sf D},{\sf V})$. The average deceleration/acceleration tuple is used to compare the similarity between each pair $({\sf D},{\sf V})$ and $({\sf D}', {\sf V}')$.

\subsection{Matrix Factorization}

The similarity matching approaches are based on domain-specific knowledge.
Collaborative filtering is a domain-free approach, relying on the identification of abstract latent factors. Matrix factorization is a popular approach of constructing latent factors, which has been implemented in recommendation system \cite{yehuda2009mftrc} and other large-scale problems \cite{gemulla2011lgmfgd}.

Consider an example of sparse matrix $R$ of $n$ pairs of $({\sf D},{\sf V})$ and $m$ road segments ${\sf R}$, as shown in Table~\ref{tab:MF}, in which each entry represents a measurement (e.g.,  ${v}$, $\vec{{d}}$ or $\vec{{a}}$). Note that some data points may be missing in $R$, denoted by ``?''.

\begin{table}[!htb]
  \centering
  \begin{tabular}{c|cccccc}
    \backslashbox{{\sf D},{\sf V}}{{\sf R}} & 1 & 2 & 3 & 4 & ... & $m$\\
    \hline
    1 & 67 & 74 & ? & 32 &  ... & 50 \\
    2 & 54 & ? & 83 & 44 &  ... & 65 \\
    \vdots & ? & 74 & 53 & ? & ... & ? \\
    $n$ & ? & 66 & 58 & ? & ... & 88 \\
  \end{tabular}
\caption{An example of sparse matrix $R$ of vehicle speed~$v$.}
\label{tab:MF}
\end{table}

The basic idea of matrix factorization is to find two low-rank ($n\times k$ and $m\times k$) matrices, $P$ and $Q$, such that $PQ^T$ can approximate $R$. Namely,
\begin{equation}
R \approx P Q^T=\hat R \label{eqn:mffc}
\end{equation}
$P$ and $Q$ can be regarded as mappings to reduce the $m,n$-dimensional space of the original dataset to a $k$-dimensional space of latent factors, where $k \ll \min(m,n)$. Denote the entry at the $i$-th column and the $j$-th row of $R$ be $r_{ij}$.

The objective of matrix factorization is find $P, Q$ such that
\begin{equation}
\min_{P,Q}\sum_{i,j}(r_{ij}-p_iq_j^T)^2+ \lambda_P||p_i||^2+\lambda_Q||q_j||^2 \label{eqn:mfobj}
\end{equation}
where $p_i$ is the $i$-th row vector of $P$, and $q_j$ is the $j$-th column vector of $Q$.
Since factorization may cause over-fitting, $\lambda_P$ and $\lambda_Q$ are used to regularize the fitting.

There are two popular approaches to compute $P, Q$ in Eqn.~(\ref{eqn:mfobj}): stochastic gradient descent \cite{gemulla2011lgmfgd} and alternating least squares \cite{yehuda2009mftrc}.
In this paper utilizes stochastic gradient descent. The basic idea is to go through all $r_{ij}$ in $R$. For each $r_{ij}$, determine the corresponding factor vectors $p_i$ and $q_j$. Then, compute the approximate value by $p_iq_j^T$ and update the parameters according to:
\begin{equation}
\begin{split}
p_i \gets p_i + \epsilon(e_{ij}q_j-\lambda_Pp_i)\\
q_j \gets q_j + \epsilon(e_{ij}p_i-\lambda_Qq_j)
\end{split}
\end{equation}
where $e_{ij}=r_{ij}-p_iq_j^T$ represents the difference between approximate value and actual value and $\epsilon$ is the learning rate.
Once $P, Q$ are determined, the estimation of a missing data $\hat{r}_{ij}$ can be estimated by $\hat{r}_{ij}=p_iq_j^T$.
All measurements (e.g.,  ${v}$, $\vec{{d}}$ or $\vec{{a}}$) can be substituted and estimated using matrix factorization. The estimated values can be utilized in the vehicle energy consumption prediction.

\subsection{Comparison with the Average}

A simple approach for estimating vehicle energy consumption is based on the global average data values (e.g., average speed) from participatory sensing data. However, each driver may deviate considerably from the average data values. To compensate for the deviations, a personalized adjustment is incorporated to improve the prediction accuracy.

Let $f_{\sf D, \sf V,\sf R}$ be a personal data value for tuple $(\sf D, \sf V)$ in environment $\sf R$, and the average data value be $\bar{f_{\sf R}}$. An adjustment function $\mathcal{D}^f_{\sf D, \sf V}(\cdot)$ is used to convert the average data value to the personal data value, such that:
\begin{equation}
f_{\sf D, \sf V,\sf R}=\mathcal{D}_f^{\sf D, \sf V} (\bar{f_{\sf R}})
\end{equation}
In this paper, a simple adjustment function is considered by the following regression model:
\begin{equation}
\mathcal{D}_f^{\sf D,V}(\bar{f_{\sf R}})=\eta_1 \bar{f_{\sf R}}^2+\eta_2 \bar{f_{\sf R}} + \eta_3 \label{eqn:adjust}
\end{equation}

\section{Empirical Evaluations} \label{sec:expval}

This section discusses the empirical evaluations of the energy consumption model and its properties.

\subsection{Setup}

The driving data from 5 drivers and 7 vehicles is collected. The information of vehicles is given in Table~\ref{tab:Vehicles}. Some drivers drove multiple vehicles, which gives totally $10$ tuples of ({\sf D},{\sf V}). Since the context of participatory sensing is considered, it suffices to consider a relatively small dataset.

\begin{table}[!htb]
\centering
  \begin{tabular}{@{}c@{\ }|@{\ }c@{\ }|@{\ }c@{\ }|@{\ }c@{\ }|@{\ }c@{\ }|@{\ }c@{}} 
\hline
\hline
    Vehicle & Maker & Model & Year & Type & Displacement\\
\hline
${\sf V}_1$ & Nissan & LEAF & 2014 & EV & NA\\
${\sf V}_2$ & Ford & Fiesta & 2013 & ICE & 1.4\\
${\sf V}_3$ & Toyota & Yaris & 2013 & ICE & 1.5 \\
${\sf V}_4$ & Hyundai & Veloster & 2014 & ICE & 1.6\\
${\sf V}_5$ & Ford & Fusion & 2012 & ICE & 2.5\\
${\sf V}_6$ & BMW & 650i & 2014 & ICE & 5.0\\
${\sf V}_7$ & Ford & F150 & 2014 & ICE & 5.0\\
\hline \hline
  \end{tabular}
  \caption{{The vehicles in the experiments.}}
  \label{tab:Vehicles}
\end{table}

\begin{figure}[H]
    \includegraphics[width=.5\textwidth]{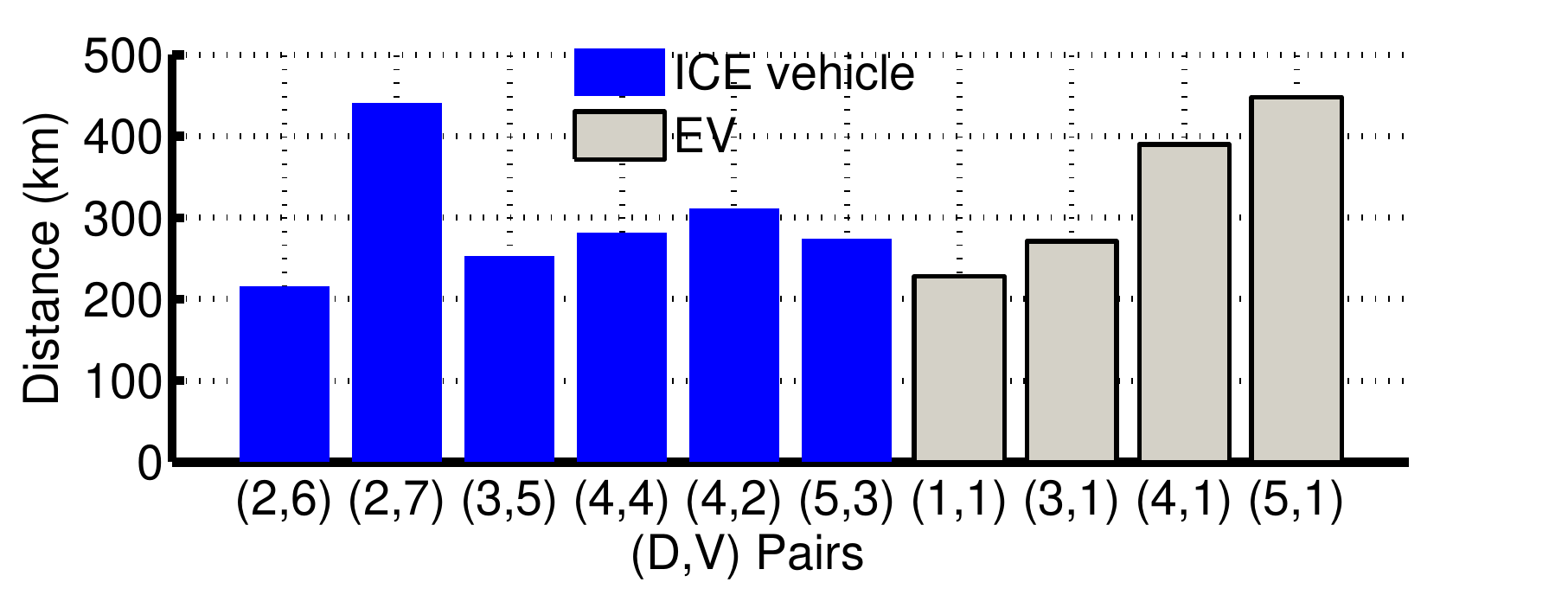}  
    \caption{Collected data of all driver-vehicle pairs.}
    \label{fig:Histdatalen}
\end{figure}

Totally 3000 km of data is collected. Fig.~\ref{fig:Histdatalen} depicts the distance of collected data for all driver-vehicle pairs. The data is then segmented into 1-km segments.

For ICE vehicles, we collected data through ELM327 devices connected to vehicles' onboard diagnostic (OBD) ports and paired with a smartphone. The collected OBD data include mass air-flow, manifold absolute pressure, intake air temperature and engine RPM. Geo-location data, accelerometer and gyroscope measurements from the smartphone are also collected.
For EVs (i.e., Nissan LEAF), high resolution state-of-charge (SOC) and vehicle speed data are collected.

\subsection{Estimation Errors of Energy Consumption Model}

The ground truth energy consumption data (i.e., ${E}_{{\sf D},{\sf V},{\sf R}}$, ${E}^{\rm mv}_{{\sf D},{\sf V},{\sf R}}$,  ${E}^{\rm id}_{{\sf D},{\sf V},{\sf R}}$) can be obtained from OBD data. From the OBD data of ICE vehicles, the fuel rate can be estimated based on mass air flow and fuel/air ratio. From the OBD data of EVs, the energy consumption is estimated by SOC and the battery capacity. Readers can refer to \cite{cmtseng2016EVextract} for the details of extraction OBD data from EVs.

\begin{figure}[!htb]
    \includegraphics[width=.48\textwidth]{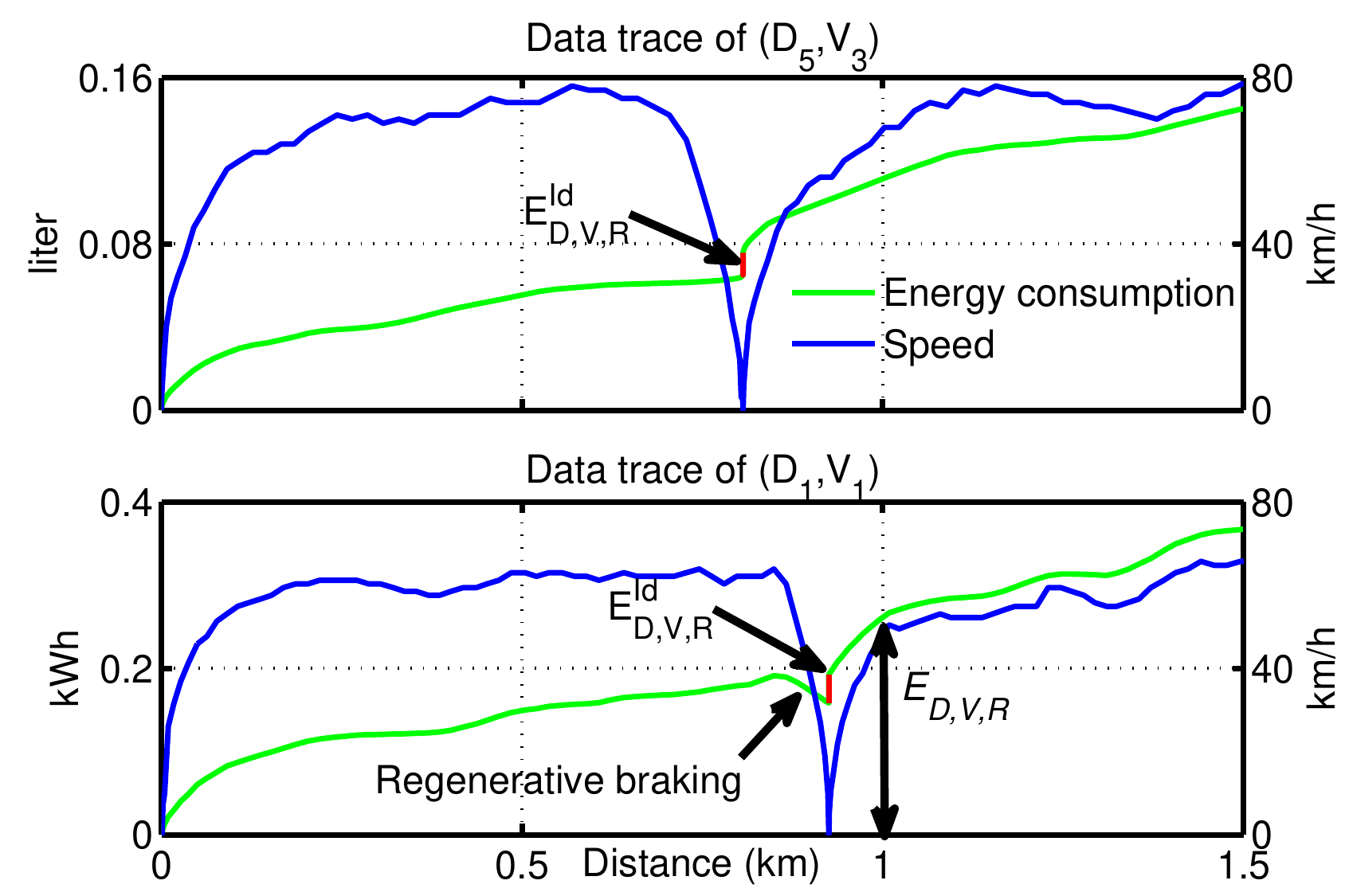}  
    \caption{Energy consumption of different driver-vehicle pairs.}
    \label{fig:groundtruthexplanation}
\end{figure}

Two metrics of error are utilized to evaluate the energy consumption predictions in this study.
The first metric of error is the per-segment error for each segment of road ${\sf R}^i$:
\begin{equation}
\varepsilon^i = \frac{( {E}^{\rm mv}_{{\sf D},{\sf V}, {\sf R}^i} + {E}^{\rm id}_{{\sf D},{\sf V}, {\sf R}^i} ) - ( \hat{E}^{\rm mv}_{{\sf D},{\sf V}, {\sf R}^i} + \hat{E}^{\rm id}_{{\sf D},{\sf V}, {\sf R}^i} )}{{E}^{\rm mv}_{{\sf D},{\sf V}, {\sf R}^i} + {E}^{\rm id}_{{\sf D},{\sf V}, {\sf R}^i}}
\label{eq:pererror}
\end{equation}
which is used to evaluate the accuracy of the energy consumption model (Eqns.~(\ref{eqn:movmodel})-(\ref{eqn:idlmodel})).
The second metric of error is the accumulative error:
\begin{equation}
\varepsilon^{\rm acc} = \frac{|{E}_{{\sf D},{\sf V}, {\sf R}} - \hat{E}_{{\sf D},{\sf V}, {\sf R}}|}{{E}_{{\sf D},{\sf V}, {\sf R}}}
\end{equation}
which is used to evaluate the energy prediction accuracy over a trip composed of many segments.
Fig.~\ref{fig:groundtruthexplanation} shows the speed profile and energy consumption of an ICE vehicle and an EV data. The idling energy consumption (${E}^{\rm id}_{{\sf D},{\sf V},{\sf R}}$) is identified from speed profile, and the moving energy consumption (${E}^{\rm mv}_{{\sf D},{\sf V},{\sf R}}$) is obtained by Eqn.~(\ref{eq:totalmodel}).

\begin{figure*}[!htb]
    \hspace{-5pt}
    \begin{subfigure}[t]{0.32\textwidth}
        \includegraphics[width=1.1\textwidth]{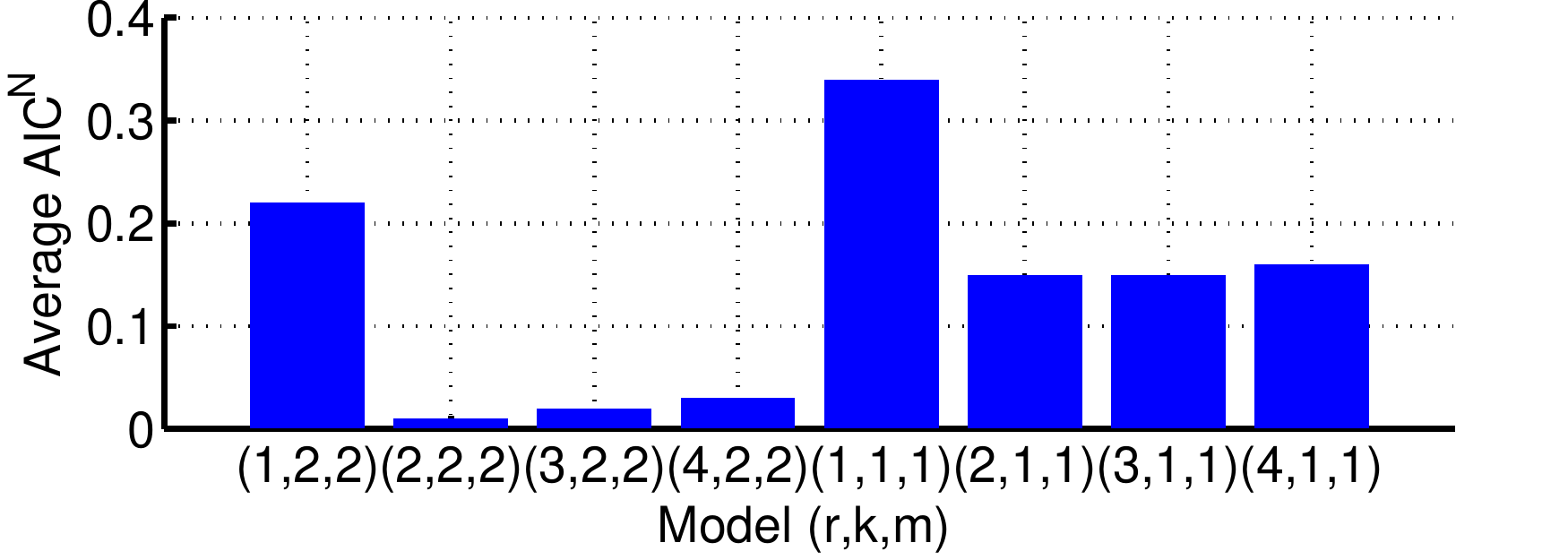} 
        \caption{Average normalized AIC values.}
        \label{fig:AIC}
    \end{subfigure}
    ~
    \begin{subfigure}[t]{0.32\textwidth}
        \includegraphics[width=1.1\textwidth]{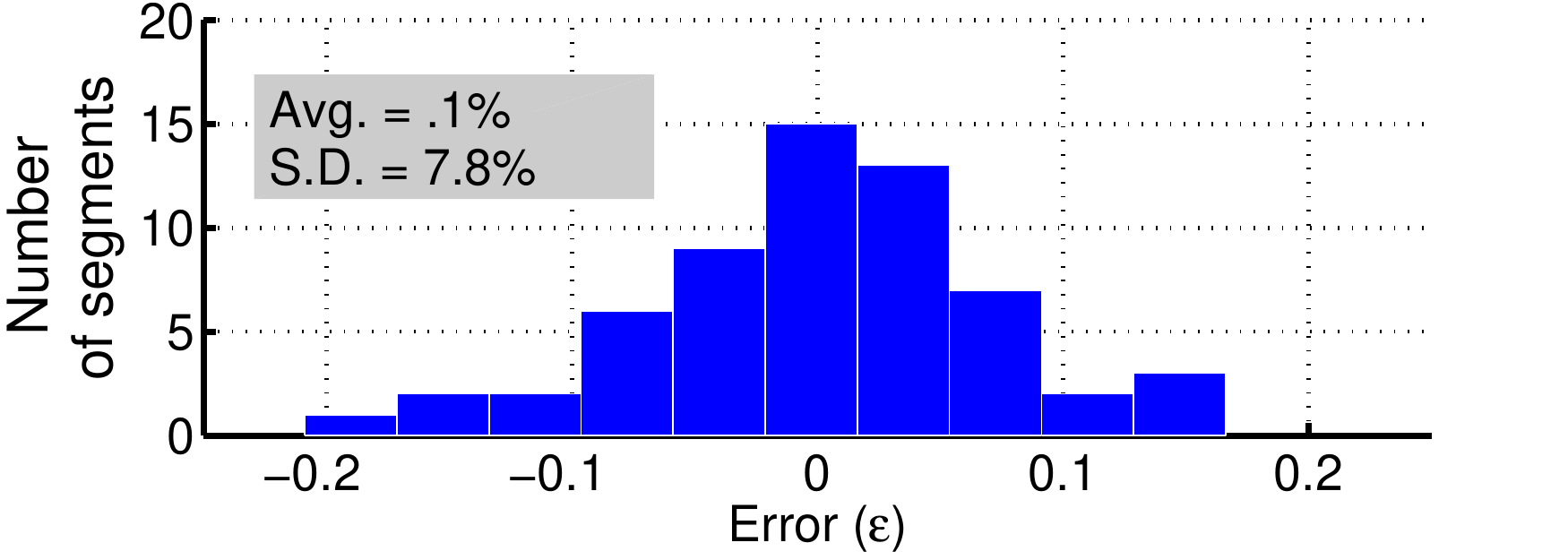} 
        \caption{{Distribution of per-segment errors.}}
        \label{fig:SegError}
    \end{subfigure}
    ~
    \begin{subfigure}[t]{0.32\textwidth}
    \includegraphics[width=1.1\textwidth]{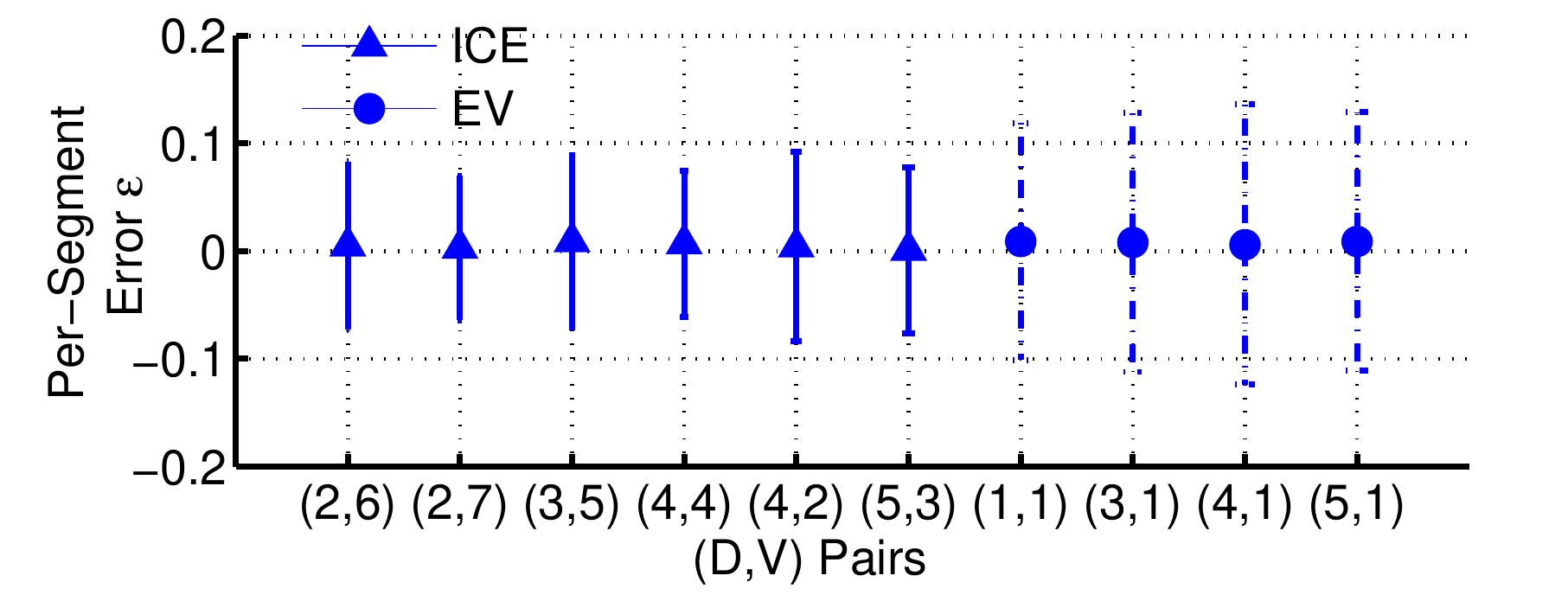} 
    \caption{Error distribution of model (2,2,2)}
    \label{fig:selfest}
    \end{subfigure}
	\caption{Experimental data and evaluation.}
\end{figure*}
\subsection{Fitness of Energy Consumption Model}

In this section, the proper powers of $v, \vec{d}, \vec{a}$ in Eqn.~(\ref{eqn:movmodel}) for model fitting are evaluated. The Akaike Information Criterion (AIC) \cite{hamparsum1987msaic} is utilized to determine the proper values of $(r,k,m)$. AIC estimates the quality of each model and balances the trade-off between the goodness of model fitting and the complexity of model. The AIC value of a model can be computed using the estimated residual in least square method. Consider the energy consumption model using $(r,k,m)$ order of powers in Eqn.~(\ref{eqn:movmodel}), the AIC value is expressed by:
\begin{equation}
{\rm {AIC}}_{(r,k,m)}=n\log\frac{\sum{\varepsilon}_i^2}{n}+2K
\end{equation}
where $\varepsilon$ is the per-segment error (see Eqn.~(\ref{eq:pererror})), $n$ is the number of segment and $K$ is the total number of estimated
regression coefficients (e.g., $r+k+m+3$). According to AIC test criterion, the smaller value makes the better model. The AIC value is normalized as ${\rm AIC}^N_{(r,k,m)}$, with respect to the smallest AIC value in all driver-vehicle pairs $(\min({{\rm AIC}}))$:
\begin{equation}
{\rm AIC}^N_{(r,k,m)}=\frac{{{\rm AIC}}_{(r,k,m)}}{\min({{\rm AIC}})}-1
\end{equation}
The mean normalized AIC values averaged over all driver-vehicle pairs for a particular $(r,k,m)$ are plotted in Fig.~\ref{fig:AIC}, which shows that $(2,2,2)$ attains the minimum, and hence, is perceived as the best setting of powers of $v, \vec{d}, \vec{a}$ in Eqn.~(\ref{eqn:movmodel}).

\subsection{Evaluation of Energy Consumption Model}
The per-segment errors of $(2,2,2)$ model for all $({\sf D},{\sf V})$ pairs are validated in this section. 80\% of collected data (called in-sample data) are randomly selected to train the regression model. The rest of data (called  out-sample data) are utilized to validate the accuracy of the model. The per-segment error distribution of out-sample data for driver-pair pair $(4,4)$ is shown in Fig.~\ref{fig:SegError}. The total number of segment is 58 (58km). The mean error is 0.1\%. The standard deviation is about 7.8\% and the distribution approaches a normal distribution. This shows that our energy consumption model is relatively accurate. The model validation results of all pairs are displayed in Fig.~\ref{fig:selfest}. Slightly higher standard deviation of per-segment error is observed for EVs, because of a lower sample rate.

\begin{figure}[!htb]
\center
\includegraphics[width=0.49\textwidth]{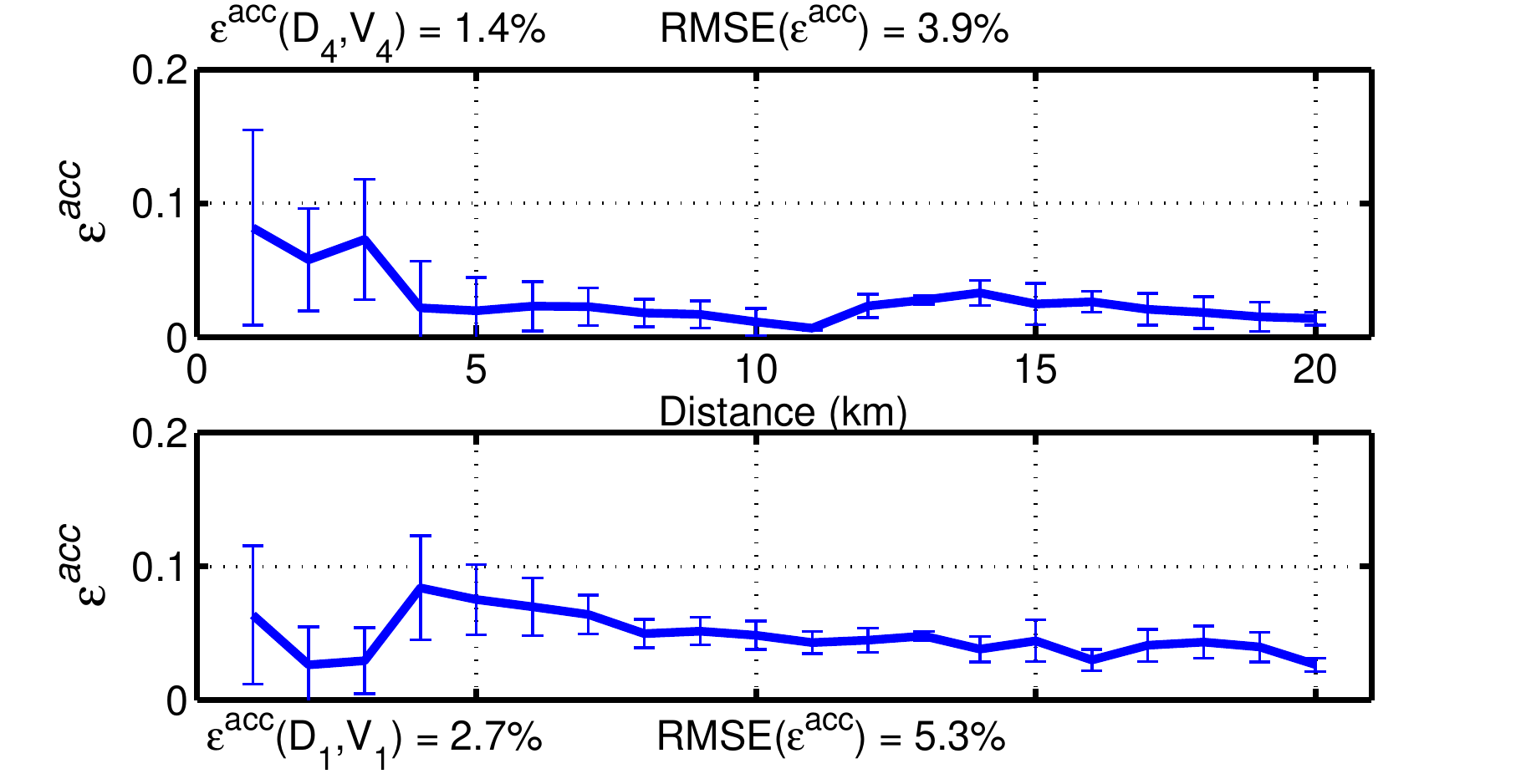} 
\caption{{Accumulative error against traveled distance.}}
\label{fig:AcuError}
\end{figure}

Since we are interested in the energy consumption of the overall trip, the accumulative error is more relevant. Fig.~\ref{fig:AcuError} shows the accumulative error against traveled distance over multiple rounds in the same route. It is observed that although the standard deviation is up to 5\%, the accumulative error is much smaller. This is due to the fact that the positive and negative deviations can offset each other over a longer distance. Therefore, the accumulative error has a lower value after a longer distance. The root mean square accumulative error RMSE($\varepsilon^{acc}$), which measures the performance over the traveled distance, is also examined. In the later case study in Sec.~\ref{sec:case}, RMSE($\varepsilon^{acc}$) will be used to evaluate the accuracy of energy consumption prediction for a designated trip.

\subsection{Dependence of Coefficients}

To properly assign the dependence of coefficients in the energy consumption model in Eqns.~(\ref{eqn:movmodel})-(\ref{eqn:idlmodel}), the distribution of coefficients between all driver-vehicle pairs is examined to identify the dependence empirically. To compare between different energy resource, the suggested conversion from the US Environmental Protection Agency (US EPA) is utilized to convert the kwh to gasoline fuel, in which 33.7 kilowatt hours of electricity is equivalent to one gallon of gasoline \cite{epa2011label}.

\begin{figure}[!htb]
\center
\includegraphics[width=0.5\textwidth]{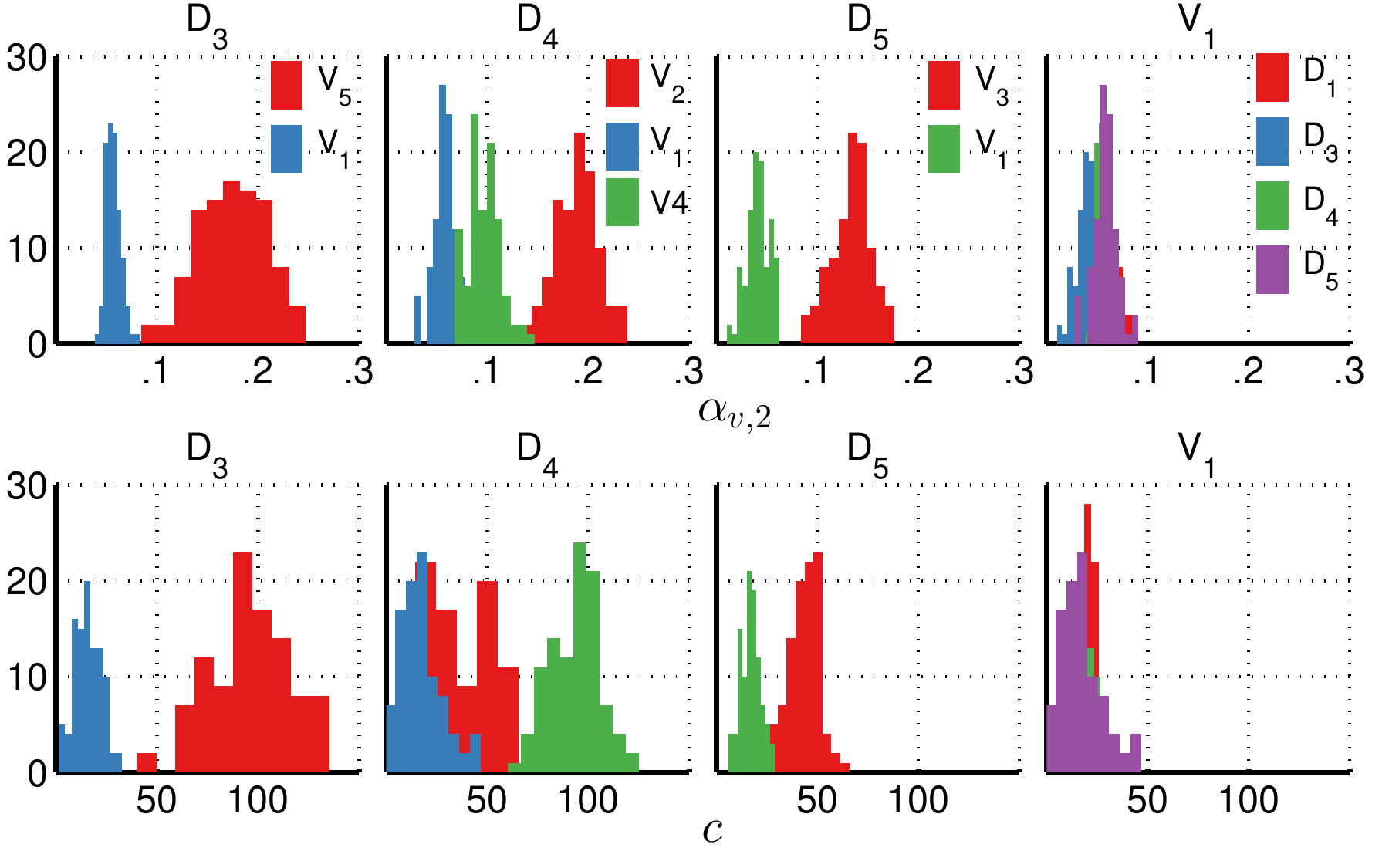} 
\caption{Distributions of coefficients $\alpha_{v,2}$ and $c$.}
\label{fig:depend}
\end{figure}

To validate the dependence, a portion (80\%) of training data is randomly drawn to train the model for each driver-vehicle pair, and the procedure is repeated 100 times to create 100 sets of coefficients for each pair. As an example, the distributions of coefficients $\alpha_{v,2}$ and $c$ to the same driver or the same vehicle are plotted in Fig.~\ref{fig:depend}. It is observed that the distributions for of coefficients $\alpha_{v,2}$ and $c$ of the same driver in different vehicles tends to be independent from another vehicle. In addition, the rightmost figures show the distributions of different drivers in the same vehicle are highly overlapping, which means the coefficients are less affected by drivers. Therefore, the coefficients $\alpha_{v,2}$ and $c$ are assigned to be vehicle dependent. The dependence of other parameters and coefficients in Table~\ref{tab:depend} are also validated.

\subsection{Driving Habits}

\begin{figure}[!htb]
        \includegraphics[width=.48\textwidth]{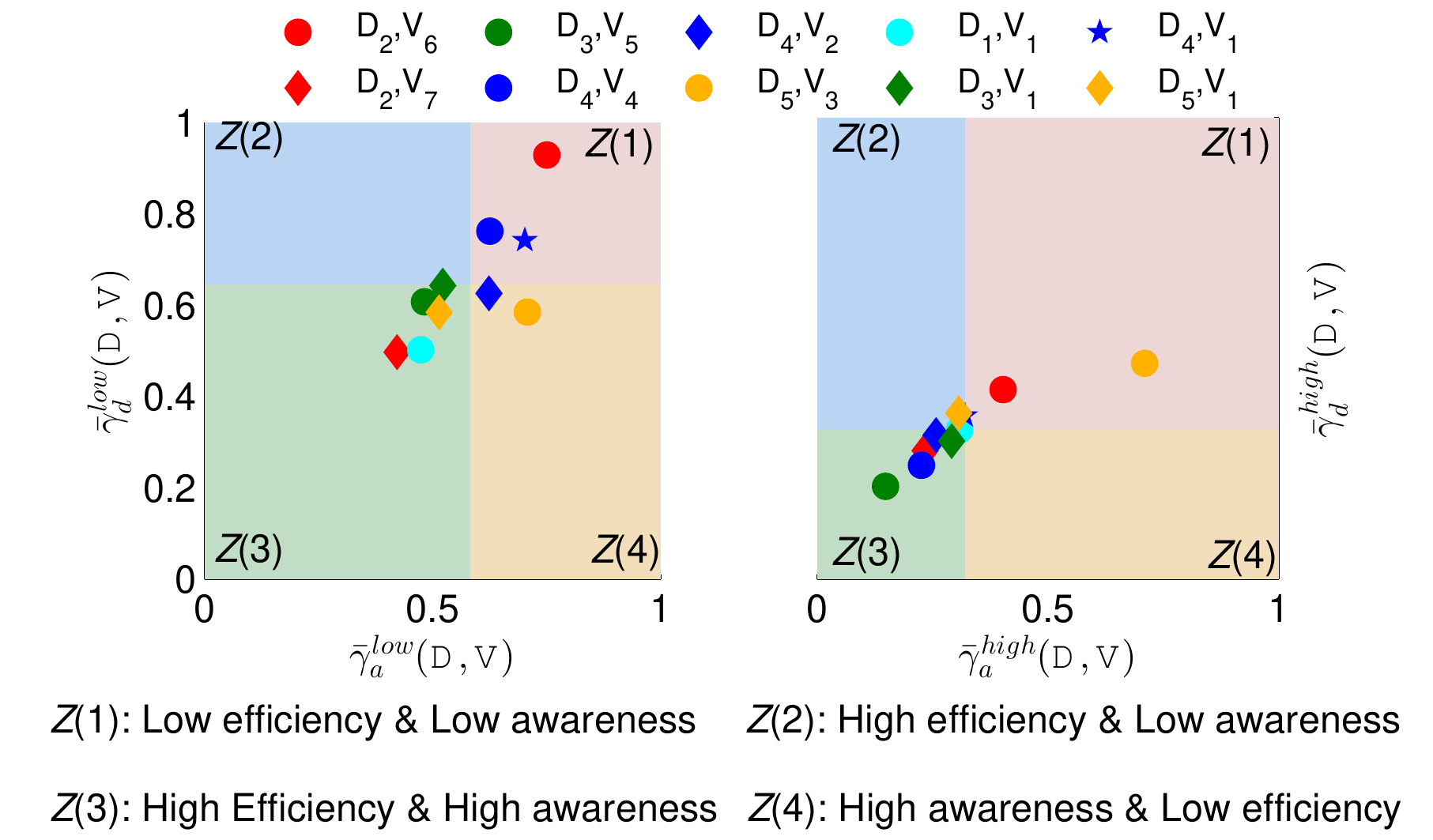}
        \caption{Regions of driving habits characterized by average accelerations/decelerations.}
        \label{fig:drvquadrant}
\end{figure}

This section compares the driving habits characterized by low-speed and high-speed average acceleration/deceleration tuple. The average accelerations/decelerations of several drivers are plotted in Fig.~\ref{fig:drvquadrant}, which are aggregated over multiple trips. Positive correlations between average acceleration and average deceleration is observed. Drivers who accelerate more tend to decelerate more. As a result, one can classify the driving habits by awareness and efficiency according to different regions in low-speed and high-speed average accelerations/decelerations plots, relative to the mean values among drivers.

\section{Case Study} \label{sec:case}

\begin{figure}[!htb]
\center
\includegraphics[width=0.49\textwidth]{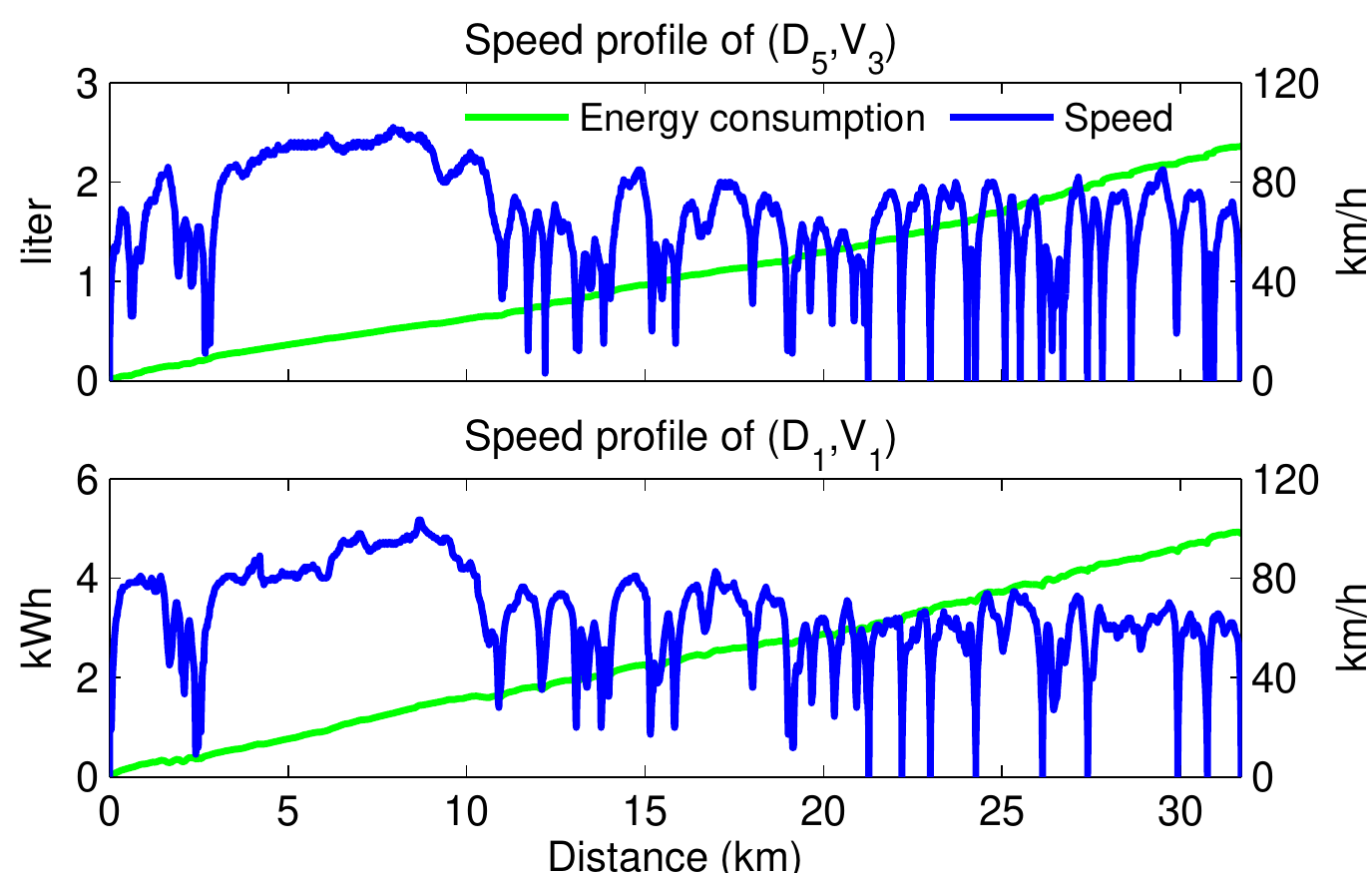} 
\caption{{Energy consumption and speed profiles of the designated route in the case study.}}
\label{fig:case_study_trace}
\end{figure}

This section presents the case study of various personalized prediction approaches. All driver-vehicle pairs are required to drive in a designated route for evaluation. The ground truth energy consumption data is also collected. Fig.~\ref{fig:case_study_trace} shows the energy consumption and speed profiles for two driver-vehicle pairs. The designated route comprises of suburban (0 to 20 km) and stop-and-go (20 to 31 km) parts. 3 rounds of driving are repeated to obtain training data and reference data.

\subsection{Personalized Vehicle Energy Consumption Prediction}

This section compares the performance of various personalized prediction approaches for vehicle energy consumption using the collected data for the designated route. The energy consumption model of each driver-vehicle pair is trained using the historical data collected from daily driving. Then the energy consumption of the route can be predicted using the collected data from different driver-vehicle pairs for the route.

For example, the energy consumption model for $\hat{E}^{\rm mv}$ of $({\sf D}_1,{\sf V}_1)$ of a particular road segment is obtained form the historic data and is given as follows.
\begin{align*}
& \hat{E}^{\rm mv} = \notag \\
& \begin{bmatrix}
-5.831\\
0.266\\
0.483\\
14.863\\
76.918\\
\end{bmatrix}^{T}
\begin{bmatrix}
{v}\\
{v}^2\\
{g} \\
{\ell} \\
1
\end{bmatrix}+\hspace{-4pt}
\begin{bmatrix}
-6.261\\
143.340\\
-139.219\\
0.034\\
-30.534\\
86.334\\
\end{bmatrix}^{T}
\begin{bmatrix}
\tau_{d}\\
\mu_{d}\\
\sigma_{d}\\
\tau_{d}^2\\
\mu_{d}^2\\
\sigma_{d}^2\\
\end{bmatrix}+\hspace{-4pt}
\begin{bmatrix}
11.067\\
11.425\\
-12.059\\
-0.056\\
-120.340\\
169.864\\
\end{bmatrix}^{T}
\begin{bmatrix}
\tau_{a}\\
\mu_{a}\\
\sigma_{a}\\
\tau_{a}^2\\
\mu_{a}^2\\
\sigma_{a}^2\\
\end{bmatrix}\label{eqn:movmodeld1v1}
\end{align*}

\begin{table*}
\begin{minipage}[!htb]{0.68\textwidth}
\hspace{-32pt}
\includegraphics[scale=0.54]{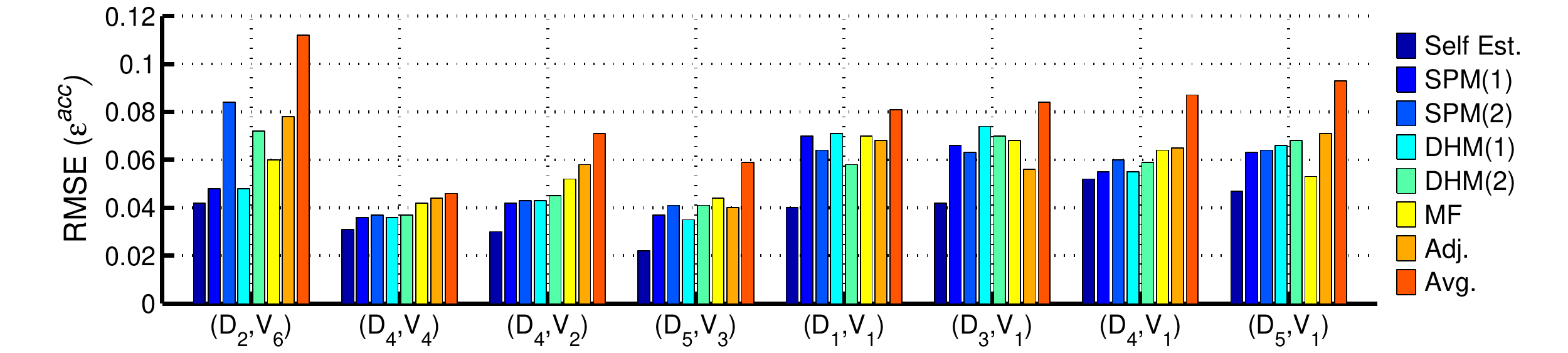}
\captionof{figure}{{Prediction error RMSE($\varepsilon^{acc}$) over all $({\sf D},{\sf V})$ pairs.}}
\label{fig:drveval}
\end{minipage}
\begin{minipage}[h]{0.32\textwidth}
  \scalebox{1}{\begin{tabular}{@{}c@{}|@{}ccc@{}}
   \hline
    \hline
    Approaches & Accuracy & Path & System \\
     &  & Matching & Complexity \\
    \hline
    SPM & High & Required & High\\
    DHM & High & No & Low\\
    MF & High & Required & Medium\\
    Adj. & High & Required & Low\\
    Avg. & Low & No & Low\\
    \hline
    \hline
  \end{tabular}
  } 	
  \caption{Summary of strengths and weaknesses of various approaches.}
  \label{tab:addisad}
\end{minipage}
\end{table*}

Various personalized prediction approaches are considered as follows:
\begin{enumerate}

\item {\em Speed Profile Matching} (SPM):
The paths among driver-vehicle pairs from historical data are matched using GPS data. The distance metric $\bar{\chi}_{({\sf D},{\sf V}),({\sf D'},{\sf V'})}$ is computed using Eqn.~(\ref{eqn:cormetric}) for all pairs of $({\sf D},{\sf V})$  and $({\sf D'},{\sf V'})$. Besides, $k$-nearest neighbors ($k$-NN) clustering is utilized to determine $k$ nearest pairs in distance metric $\bar{\chi}_{({\sf D},{\sf V}),({\sf D'},{\sf V'})}$. The similarity matching approach based on $k$ nearest pairs is denoted by SPM($k$)

\item {\em Driving Habit Matching} (DHM):
The low-speed and high-speed average deceleration/acceleration tuple $\big(\bar{\gamma}_{a}^{\rm low}({\sf D},{\sf V}),\bar{\gamma}_{a}^{\rm high}({\sf D},{\sf V}),\bar{\gamma}_{d}^{\rm low}({\sf D},{\sf V}),$$\bar{\gamma}_{d}^{\rm high}({\sf D},{\sf V})\big)$ are computed for every driver-vehicle pair $({\sf D},{\sf V})$. The low-speed and high-speed average deceleration/acceleration tuple defines a 4-dimensional data space. The similarity matching approach based on $k$ nearest pairs in the 4-dimensional data space is denoted by DHM($k$).

\item {\em Matrix Factorization} (MF):
The paths among driver-vehicle pairs from collected data using GPS data are matched before employing matrix factorization. The matrix factorization approach is denoted by MF.

\item {\em Average Data Values} (Avg.):
Using only the average data values are used (e.g., average speed) for vehicle energy consumption prediction. Sometimes, the average speed is observed to be very close to the speed limit. The average data based approach is denoted by Avg.

\item {\em Adjusted Personal Data Values} (Adj.):
The adjustment function in Eqn.~(\ref{eqn:adjust}) is used to convert the average data values to the personal data values. The adjusted personal data based approach is denoted by Adj.

\item {\em Self-Estimation} (Self Est.):
Using only one's own data in energy consumption model of the same road segment in Eqns.~(\ref{eqn:movmodel})-(\ref{eqn:idlmodel}) is also considered. The self-estimation approach is denoted by Self Est. Self-estimation is a benchmark, which essentially validates the accuracy of the energy consumption model without using the data of other driver-vehicle pairs. In practice, one's own data of the same road segment may not be always present, as the driver has not traveled such a route before.

\end{enumerate}

Fig.~\ref{fig:drveval} compares the prediction errors in terms of RMSE, against the ground truth energy consumption over all driver-vehicle pairs.
The strengths and weaknesses of each prediction approach are summarized in Table~\ref{tab:addisad}.

Self Est. is observed to have $5\%$ error, which is the lowest among all approaches, because one's own driving data on the same route is the most accurate source for prediction, in spite of the presence of different traffic condition. SPM is observed to have a close prediction error with Self Est.
Avg. is observed to have the largest error, because of the considerable deviation from individual drivers from the average. Adj. can improve the accuracy of comparison with the average.
Notably, DHM is observed to perform relatively well, even though it does not require path matching using GPS data. Therefore, driving habits are a good indicator of vehicle energy consumption. In summary, DHM provides good accuracy without GPS data, which has low complexity for system implementation.

\subsection{Distance-to-Empty Prediction for EV}

In this section, our approaches are applied to the application of DTE prediction for EV (i.e., Nissan LEAF ) using other ICE vehicle data. For the convenience of comparison, certain routes are selected and all drivers to required to travel the same route at least $3$ times for evaluations.

The data collected from Nissan LEAF includes:
\begin{enumerate}

\item State-of-charge (SOC), denoted by $\mathcal{S}$, which indicates the remaining battery level.

\item Initial capacity of the battery, denoted by $\mathcal{B}_A$. 

\item Battery pack voltage when driving, denoted by $\mathcal{B}_V$.

\end{enumerate}
The remaining energy ($\Delta{\mathcal{E}}^t$) in battery at time $t$ is given by:
\begin{equation}
    \Delta{\mathcal{E}}^t=\mathcal{S}^t\times \mathcal{B}_A\times \mathcal{B}_V
    \label{eq:batteryremaining}
\end{equation}
If the future average power intensity ($\bar{\mathcal{P}}$) is known, then estimated DTE is given by:
\begin{equation}
    {\rm \widehat{DTE}}=\dfrac{\Delta{\mathcal{E}^t}}{\bar{\mathcal{P}}}
    \label{eq:DTEequation}
\end{equation}

The DTE prediction based on approaches using participatory sensing data is compared with the on-board DTE meter on Nissan LEAF (also known as Guess-O-Meter), which is captured by a camera mounted over the dashboard. To compare the effectiveness of DTE prediction, the deviation between the true DTE (which is computed in an offline manner) and the estimated ${\rm \widehat{DTE}}$ is measured by:
\begin{equation}
    {\Delta{\rm DTE}}=\rm DTE-{\rm \widehat{DTE}}
    \label{eq:DTEequation2}
\end{equation}
The deviation between true DTE and the on-board DTE meter on Nissan LEAF is compared.

\begin{figure}[!htb]
\center
    \includegraphics[width=.5\textwidth]{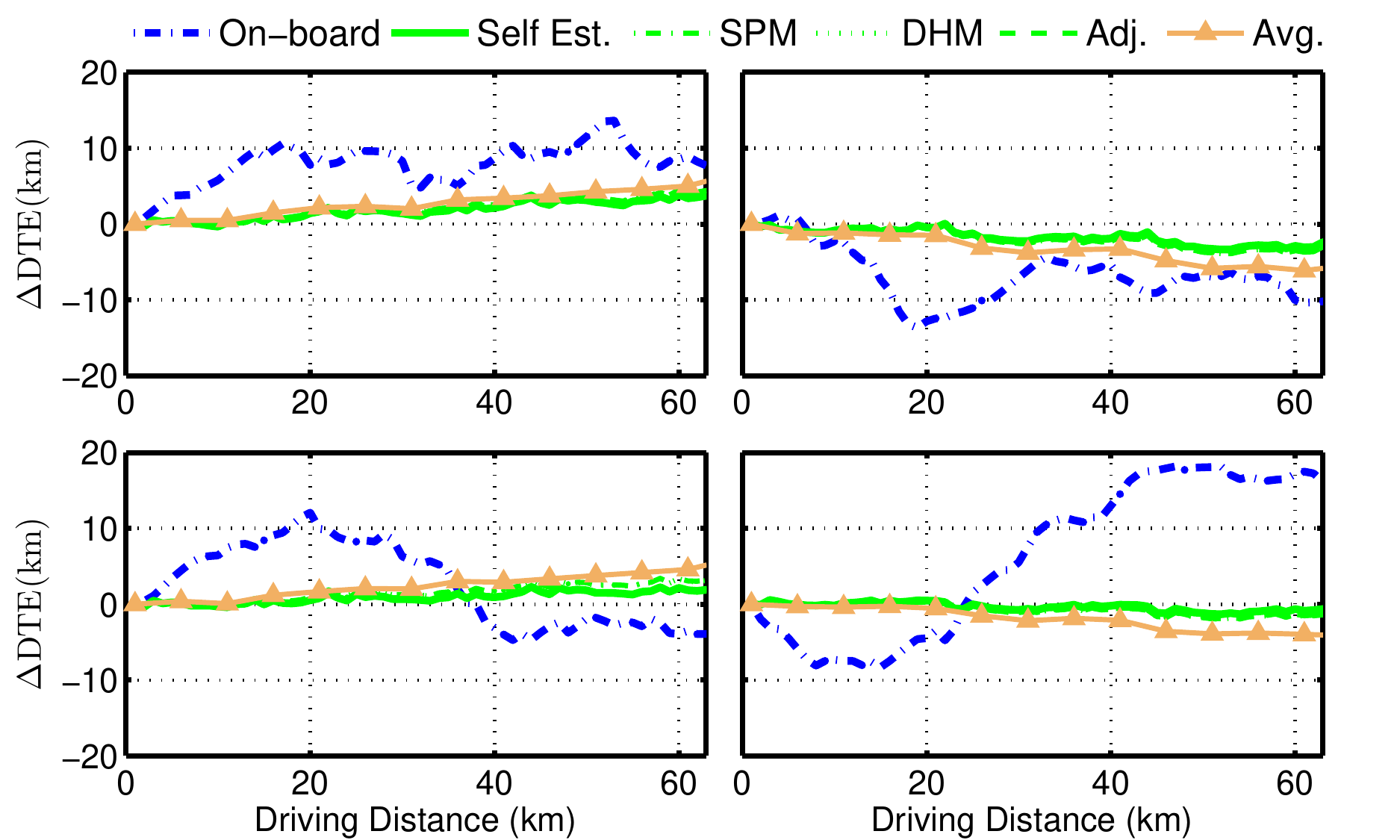}
\caption{Deviations of DTE prediction for various approaches.}
\label{fig:DTE}
\end{figure}

The results are plotted in Fig.~\ref{fig:DTE} for four different trips. It is observed that all approaches using participatory sensing data can significantly outperform the one provided by on-board DTE provided by the on-board DTE meter on Nissan LEAF. Notably, SPM, DHM and Adj. perform very close to benchmark Self Est., whereas Avg. gives relatively inferior performance.  In summary, DHM consistently provides good accuracy with low system complexity.

\section{Conclusion} \label{sec:concl}

In this paper, various methodologies of utilizing participatory sensing data for personalized prediction of vehicle energy consumption were investigated. Several approaches were studied and compared, including: (1) comparison with the average using personalized adjustment, (2) two similarity matching approaches based on driver/vehicle/environment-dependent factors using speed profile matching and driving habit matching, and (3) a collaborative filtering approach that uses matrix factorization. Our empirical evaluations show that participatory sensing data can significantly improve prediction accuracy. Among all approaches, similarity matching approach based on driving habits provides good accuracy (as compared to a benchmark of self-estimation using one's own driving data) with low system complexity. To evaluate the effectiveness, a case study of DTE prediction for EVs is conducted based on the participatory sensing data. In summary, similarity matching based on driving habit can provide a practical solution of DTE prediction for EVs, which significantly outperforms the on-board DTE meter on Nissan LEAF.

Despite the promising results given by our study, there are several practical issues and limitations to be recognized:
\begin{itemize}

\item {\em  Road Grade}: This is the elevation of roads. There are public mapping APIs to provide road elevation data. This can be added to the future energy consumption model.

\item {\em Weather and Traffic}: Our study assumes mild weather and traffic conditions. But our energy consumption model can be extended to incorporate additional parameters to capture the impacts of weather in the vehicle model (e.g., weather types and route conditions). The vehicle speed from participatory sensing data naturally reflect the traffic condition to a certain extent, however, the data would need to be updated more frequently. Wind speed and road surface conditions also affect vehicle energy consumption, but are more difficult to measure.

\item {\em More Vehicle Information}: The weight of vehicle and tire pressures of vehicle would also introduce the error to the system, the error can be minimized obtaining more data from vehicle API (e.g., tire pressure indicator).

\item {\em Change of Vehicle State}: The engine/gearbox efficiencies or battery efficiencies change over time, due to aging of vehicle or system upgrades. Hence, the energy consumption model should adapt to these changes using new data to update the coefficients.

\end{itemize}

A study is planned to be conducted in future work to provide more comprehensive insights in diverse practical settings, addressing the preceding issues.

\bibliographystyle{IEEEtran}
\bibliography{ref}

\begin{thebibliography}{10}
\providecommand{\url}[1]{#1}
\csname url@samestyle\endcsname
\providecommand{\newblock}{\relax}
\providecommand{\bibinfo}[2]{#2}
\providecommand{\BIBentrySTDinterwordspacing}{\spaceskip=0pt\relax}
\providecommand{\BIBentryALTinterwordstretchfactor}{4}
\providecommand{\BIBentryALTinterwordspacing}{\spaceskip=\fontdimen2\font plus
\BIBentryALTinterwordstretchfactor\fontdimen3\font minus
  \fontdimen4\font\relax}
\providecommand{\BIBforeignlanguage}[2]{{%
\expandafter\ifx\csname l@#1\endcsname\relax
\typeout{** WARNING: IEEEtran.bst: No hyphenation pattern has been}%
\typeout{** loaded for the language `#1'. Using the pattern for}%
\typeout{** the default language instead.}%
\else
\language=\csname l@#1\endcsname
\fi
#2}}
\providecommand{\BIBdecl}{\relax}
\BIBdecl

\bibitem{campbell2008parsense}
A.~T. Campbell, S.~B. Eisenman, N.~D. Lane, E.~Miluzzo, R.~A. Peterson, H.~Lu,
  X.~Zheng, M.~Musolesi, K.~Fodor, and G.-S. Ahn, ``The rise of people-centric
  sensing,'' \emph{IEEE Internet Comput.}, vol.~12, no.~4, pp. 12--21, 2008.

\bibitem{ganti2010greengps}
R.~K. Ganti, N.~Pham, H.~Ahmadi, S.~Nangia, and T.~F. Abdelzaher, ``{GreenGPS}:
  a participatory sensing fuel-efficient maps application,'' in \emph{ACM
  Mobile Systems, Applications, and Services ({Mobisys})}, 2010.

\bibitem{stefan2013modularecev}
S.~Grubwinkler and M.~Lienkamp, ``A modular and dynamic approach to predict the
  energy consumption of electric vehicles,'' in \emph{Conf. Future Automotive
  Technology}, 2013.

\bibitem{gemulla2011lgmfgd}
R.~Gemulla, P.~J. Haas, E.~Nijkamp, and Y.~Sismanis, ``Large-scale matrix
  factorization with distributed stochastic gradient descent,'' in \emph{ACM
  Conf. Knowledge Discovery and Data Mining ({SIGKDD})}, 2011.

\bibitem{yehuda2009mftrc}
Y.~Koren, R.~Bell, and C.~Volinsky, ``Matrix factorization techniques for
  recommender systems,'' \emph{J. of Computer}, vol.~42, pp. 30--37, 2009.

\bibitem{karin2012eeroutealg}
K.~Kraschl-Hirschmann and M.~Fellendorf, ``Estimating energy consumption for
  routing algorithms,'' in \emph{IEEE Intelligent Vehicles Symp.}, 2012.

\bibitem{eugene2013rtbattery}
E.~Kim, J.~Lee, and K.~G. Shin, ``Real-time prediction of battery power
  requirements for electric vehicles,'' in \emph{IEEE/ACM Int. Conf. on
  Cyber-Physical Systems}, 2013.

\bibitem{alesaandn2002statmdlfc}
A.~Cappiello, I.~Chabini, E.~K. Nam, A.~Lue, and M.~A. Zed, ``A statistical
  model of vehicle emissions and fuel consumption,'' in \emph{IEEE Intelligent
  Transportation Systems Conf.}, 2002.

\bibitem{javier2013mbasedrdr}
J.~A. Oliva, C.~Weihrauch, and T.~Bertram, ``A model-based approach for
  predicting the remaining driving range in electric vehicles,'' in \emph{IEEE
  Prognostics and Health Management}, 2013.

\bibitem{qichiyang2011arterialmdltraj}
Q.~Yang, K.~Boriboonsomsin, and M.~Barth, ``Arterial roadway energy/emissions
  estimation using modal-based trajectory reconstruction,'' in \emph{IEEE Int.
  Transportation Systems Conf.}, 2011.

\bibitem{cloudthink15}
E.~Wilhelm, J.~Siegel, S.~Mayer, L.~Sadamori, S.~Dsouza, C.-K. Chau, and
  S.~Sarma, ``Cloudthink: A scalable secure platform for mirroring
  transportation systems in the cloud,'' \emph{Transport}, vol.~30, no.~3,
  2015.

\bibitem{dornbush2007streetsmart}
S.~Dornbush and A.~Joshi, ``Streetsmart traffic: Discovering and disseminating
  automobile congestion using {VANET},'' in \emph{IEEE Vehicular Technology
  Conf.}, 2007.

\bibitem{cmtseng2015pardte}
C.-M. Tseng, C.-K. Chau, S.~Dsouza, and E.~Wilhelm, ``A participatory sensing
  approach for personalized distance-to-empty prediction and green
  telematics,'' in \emph{ACM Int. Conf. Future Energy Systems ({e-Energy})},
  2015.

\bibitem{cmtseng2014social}
C.-M. Tseng, S.~Dsouza, and C.-K. Chau, ``A social approach for predicting
  distance-to-empty in vehicles,'' in \emph{ACM Int. Conf. Future Energy
  Systems ({e-Energy})}, 2014.

\bibitem{burke1994electronic}
M.~J. Burke, N.~Sarafopoulos, and V.~Q. To, ``Electronic system and method for
  calculating distance to empty for motorized vehicles,'' 1994, {US} Patent
  5,301,113.

\bibitem{rodgers2013conventional}
L.~Rodgers, E.~Wilhelm, and D.~Frey, ``Conventional and novel methods for
  estimating an electric vehicle's distance to empty,'' in \emph{ASME Intl.
  Conf. Advanced Vehicle Technologies}, 2013.

\bibitem{anastasia2014onpevreg}
A.~Bolovinou, I.~Bakas, A.~Amditis, F.~Mastrandrea, and W.~Vinciotti, ``Online
  prediction of an electric vehicle remaining range based on regression
  analysis,'' in \emph{IEEE Int. Electric Vehicle Conf.}, 2014.

\bibitem{haiyu2012drvpatternev}
H.~Yu, F.~Tseng., and R.~McGee, ``Driving pattern identification for ev range
  estimation,'' in \emph{IEEE Int.l Electric Vehicle Conf.}, 2012.

\bibitem{kanok2012ecorttraffic}
K.~Boriboonsomsin, M.~J. Barth, W.~Zhu, and A.~Vu, ``Eco-routing navigation
  system based on multisource historical and real-time traffic information,''
  \emph{IEEE Trans. Intell. Transp. Syst.}, vol.~13, no.~4, pp. 1694--1704,
  2012.

\bibitem{martin2011eoptev}
M.~Sachenbacher, M.~Leucker, A.~Artmeier, and J.~Haselmayr, ``Efficient
  energy-optimal routing for electric vehicles,'' in \emph{AAAI Conf.
  Artificial Intelligence}, 2011.

\bibitem{stefan2013syscloud}
S.~Grubwinkler, M.~Kugler, and M.~Lienkamp, ``A system for cloud-based
  deviation prediction of propulsion energy consumption for evs,'' in
  \emph{IEEE Int. Conf. Vehicular Electronics and Safety}, 2013.

\bibitem{cmtseng2016privacy}
C.-M. Tseng and C.-K. Chau, ``On the privacy of crowd-sourced data collection
  for distance-to-empty prediction and eco-routing,'' in \emph{ACM Workshop on
  Electric Vehicle Systems, Data and Applications ({EV-Sys})}, 2016.

\bibitem{ckchau2016phevopt}
C.-K. Chau, K.~M. Elbassioni, and C.-M. Tseng, ``Fuel minimization of plug-in
  hybrid electric vehicles by optimizing drive mode selection,'' in \emph{ACM
  Int. Conf. Future Energy Systems ({e-Energy})}, 2016.

\bibitem{ckt2016dm}
C.-K. Chau, K.~Elbassioni, and C.-M. Tseng, ``Drive mode optimization and path
  planning for plug-in hybrid electric vehicles,'' \emph{to appear in IEEE
  Trans. Intell. Transp. Syst.}, 2017.

\bibitem{eva2001idrvpattern}
E.~Ericsson, ``Indepentent driving pattern factors and their influence on
  fuel-use and exhaust emission factor,'' \emph{J. of Transportation Research},
  vol.~6, no.~5, pp. 325--345, 2001.

\bibitem{Stan2007DTW}
S.~Salvador and P.~Chan, ``Toward accurate dynamic time warping in linear time
  and space,'' \emph{J. of Intelligent Data Analysis}, vol.~11, no.~5, pp.
  561--580, 2007.

\bibitem{cmtseng2016EVextract}
C.-M. Tseng, W.~Zhou, M.~A. Hashmi, C.-K. Chau, S.~G. Song, and E.~Wilhelm,
  ``Data extraction from electric vehicles through {OBD} and application of
  carbon footprint evaluation,'' in \emph{ACM Workshop on Electric Vehicle
  Systems, Data and Applications ({EV-Sys})}, 2016.

\bibitem{hamparsum1987msaic}
H.~Bozdogan, ``Model selection and akaike's information criterion (aic): The
  general theory and its analytical extensions,'' \emph{J. of Psychometrika},
  vol.~52, pp. 345--370, 1987.

\bibitem{epa2011label}
U.~E.~P. Agency, \emph{New Fuel Economy and Environment Labels for a New
  Generation of Vehicles}.\hskip 1em plus 0.5em minus 0.4em\relax US
  Environmental Protection Agency, 2011.

\end{thebibliography}

\end{document}